\documentclass[aps,prd,superscriptaddress,nofootinbib,twocolumn]{revtex4-2}

\usepackage{mathtools}
\usepackage{amsfonts}
\usepackage{mathrsfs}
\usepackage{amssymb}

\usepackage{bbm}
\usepackage{slashed}
\usepackage{amsmath}
\usepackage{bm}
\usepackage{tensor}

\usepackage{graphicx}
\usepackage{color}
\usepackage{colortbl}
\usepackage{array}
\usepackage[dvipsnames]{xcolor}

\usepackage{float}
\usepackage{placeins}
\usepackage{booktabs}
\usepackage[caption=false]{subfig}
\usepackage{makecell}
\usepackage{tabstackengine}

\usepackage{xspace}
\usepackage{siunitx}
\usepackage{hyperref}
\usepackage[nameinlink]{cleveref}
\usepackage{bookmark}

\usepackage{xifthen}
\usepackage{xcolor}
\usepackage{enumitem}
\hypersetup{
	colorlinks,
	linkcolor={red!75!black},
	citecolor={blue!75!black},
	urlcolor={blue!75!black}
}

\usepackage[utf8]{inputenc}

\setkeys{Gin}{width=0.48\textwidth}

\newcommand*{\eg}{e.g.\@\xspace}
\newcommand*{\ie}{i.e.\@\xspace}

\makeatletter
\newsavebox\myboxA
\newsavebox\myboxB
\newlength\mylenA

\newcommand*\xoverline[2][0.75]{%
	\sbox{\myboxA}{$\m@th#2$}%
	\setbox\myboxB\null
	\ht\myboxB=\ht\myboxA%
	\dp\myboxB=\dp\myboxA%
	\wd\myboxB=#1\wd\myboxA
	\sbox\myboxB{$\m@th\overline{\copy\myboxB}$}
	\setlength\mylenA{\the\wd\myboxA}
	\addtolength\mylenA{-\the\wd\myboxB}%
	\ifdim\wd\myboxB<\wd\myboxA%
	\rlap{\hskip 0.5\mylenA\usebox\myboxB}{\usebox\myboxA}%
	\else
	\hskip -0.5\mylenA\rlap{\usebox\myboxA}{\hskip 0.5\mylenA\usebox\myboxB}%
	\fi}
\makeatother

\graphicspath{{./figures/}}

\newcommand{\gettitle}{Spectral properties and observables in ultracold Fermi gases}

\newcommand{\getHeidelbergAffiliation}{\affiliation{Institut f\"ur Theoretische Physik, Universit\"at Heidelberg, Philosophenweg 16, 69120 Heidelberg, Germany}}
\newcommand{\getEMMIAffiliation}{\affiliation{ExtreMe Matter Institute EMMI, GSI, Planckstr. 1, 64291 Darmstadt, Germany}}

\hypersetup{
	colorlinks,
	linkcolor={red!75!black},
	citecolor={blue!75!black},
	urlcolor={blue!75!black},
	pdftitle={\gettitle},
	pdfauthor={Dizer, Horak, Pawlowski},
	pdfkeywords={analytic continuation}
	{correlations functions} {ultracold fermi gases}
	{functional renormalisation group}
	{real time} {spectral function} 
	bookmarksopen=true,
	bookmarksopenlevel=2,
	bookmarksnumbered=true
}

\allowdisplaybreaks

\begin{document}

\title{\gettitle}

\author{Eugen Dizer}
\getHeidelbergAffiliation

\author{Jan Horak}
\getHeidelbergAffiliation

\author{Jan M. Pawlowski}
\getHeidelbergAffiliation
\getEMMIAffiliation

\begin{abstract}

We calculate non-perturbative self-consistent fermionic and bosonic spectral functions of ultracold Fermi gases with the  spectral functional approach. This approach allows for a direct real-time computation of non-perturbative correlation functions, and in the present work we use spectral Dyson-Schwinger equations. We focus on the normal phase of the spin-balanced Fermi gas and provide numerical results for the full fermionic and bosonic spectral functions. The spectral functions are then used for the determination of the equation of state, the Tan contact and ejection rf spectra at unitarity. These results are compared to experimental data, the self-consistent T-matrix approach and lattice results. 
Our approach offers a wide range of applications, including the ab initio calculation of transport and spectral properties of the superfluid phase in the BCS-BEC crossover. 

\end{abstract}

\maketitle

\section{Introduction} \label{sec:Introduction}

Ultracold Fermi gases~\cite{Bloch:2008zzb,Giorgini:2008zz,Zwerger2016} represent a versatile tool for studying strongly correlated quantum systems at low energies. The experimental control over the system parameters allows for the investigation of a wide range of phenomena, such as the BCS-BEC crossover~\cite{greiner2003emergence,Chin_2004,Perali:2004zza,PhysRevLett.122.203402,Strinati:2018wdg}, the formation of polarons~\cite{PhysRevLett.102.230402,PhysRevLett.103.170402,massignan2014polarons,Scazza:2022bez}, and the emergence of superfluidity~\cite{zwierlein2005vortices,Zwierlein:2006zz,Ku2012}. Various important properties of the gas, such as transport and scattering properties~\cite{Taylor:2010ju,Enss:2010qh,Frank:2020oef} or the excitation spectrum~\cite{Greiner:2005zz,PhysRevB.85.024517}, are encoded in its spectral functions. 

Strong correlations, e.g.~close to unitarity~\cite{Bloch:2008zzb}, render the theoretical description of interacting Fermi gases a non-perturbative problem. Therefore, various non-perturbative approaches, such as Quantum Monte Carlo (QMC) simulations~\cite{Burovski:2006zz,Magierski:2008wa}, self-consistent T-matrix theory~\cite{Haussmann:2007zz, PhysRevA.80.063612,PhysRevA.77.061605,PhysRevA.81.023622,Hanai2014} or Dyson-Schwinger equation (DSE)~\cite{PhysRevA.77.023626, Diehl:2005an, Boettcher:2012dh} and functional renormalisation group (fRG)~\cite{Diehl:2007th, Diehl:2007ri, Diehl:2009ma} approaches, have been employed to describe the system.

The computation of spectral functions requires access to the fermionic and bosonic self-energies at real frequencies. While QMC simulations are formulated at imaginary frequencies by construction, also functional methods such as DSE and fRG are usually formulated in Euclidean spacetime due to significantly reduced computational costs. However, this requires an analytic continuation of the correlation functions from Matsubara to real frequencies, which is an ill-conditioned numerical task accompanied by large systematic uncertainties~\cite{Jarrell1996}.

Recently, the \textit{spectral functional approach} has been put forward~\cite{Horak:2020eng}. Notably, this approach allows for the direct computation of non-perturbative correlation functions at real frequencies: with the spectral representations of propagators and vertices in non-perturbative diagrams, the respective frequency sums or integrals can be performed analytically, and the remaining real spectral integrals are readily computed numerically. The spectral functional approach has been developed within the framework of Dyson-Schwinger equations, but it applies to all functional approaches. The spectral functional renormalisation has been set up and used in \cite{Fehre:2021eob, Braun:2022mgx}, and the spectral approach is tailor made for application in 2PI resummation schemes~\cite{Berges:2004yj}, where all correlation functions are constructed from non-perturbative propagators. It has been successfully used for the non-perturbative computation of spectral functions and bound state properties in scalar theories~\cite{Horak:2020eng, Horak:2023hkp, Eichmann:2023tjk}, gauge theories~\cite{Horak:2022aza, Horak:2022myj, Horak:2021pfr, Horak:2023rzx}, and quantum gravity~\cite{Fehre:2021eob}. 

In this work,  we use the spectral functional approach for the computation of non-perturbative self-consistent single-particle spectral functions in the normal phase of a strongly interacting spin-balanced Fermi gas for all scattering lengths $a$, and in particular close to a Feshbach resonance~\cite{RevModPhys.82.1225} with $a\to\infty$. This constitutes the first application of the spectral functional approach to non-relativistic systems. Specifically, we set up spectral Dyson-Schwinger equations for the fermionic and bosonic self-energies at real frequencies, yielding direct access to their spectral functions. With the spectral functions we also determine the density equation of state~\cite{Ku2012}, the Tan contact~\cite{Rossi:2013tik} and ejection radio-frequency (rf) spectra~\cite{PhysRevLett.122.203402} at unitarity. Our results agree with results from the self-consistent T-matrix approximation~\cite{Haussmann:2007zz, PhysRevA.80.063612}, but the flexibility of the spectral functional approaches gives access to more generic expansion schemes. 

The present paper is organised as follows. In \Cref{sec:microscopic-model}, we introduce our microscopic model for the spin-balanced Fermi gas. In \Cref{sec:spectral_DSE}, we set up the spectral Dyson-Schwinger equations for bosonic and fermionic propagators. Our numerical results for the spectral functions as well as the density, rf spectra and Tan contact are presented in \Cref{sec:results}. We conclude in \Cref{sec:Conclusion}.

\section{Microscopic Model} \label{sec:microscopic-model}

We consider a non-relativistic two-component Fermi gas described by the Euclidean classical action
\begin{align} \nonumber 
    S[\psi] = \int_{0}^{\beta} d\tau \int d^3 x\, \Big[ \sum_{\sigma=\uparrow,\downarrow} \psi_{\sigma}^* (&\partial_{\tau} - \nabla^2 - \mu_{\sigma}) \psi_{\sigma}  \\[1ex]
    &+ \lambda\, \psi_{\uparrow}^*\psi_{\downarrow}^*\psi_{\downarrow}\psi_{\uparrow} \Big] \,,
\label{eq:microscopic-action}
\end{align}
where $\lambda$ is the coupling constant of the contact interaction. If used on the mean field level, $\lambda$ would be related directly to the scattering length $a$ via \cite{Diehl:2009ma} 
\begin{align}
	\lambda = 8 \pi a\,.  
\label{eq:lambda-a}
\end{align}
Instead of using the common split into bare coupling and cutoff contribution, \labelcref{eq:lambda-a} is readily converted into a cutoff-independent relation including quantum effects, for more details see \Cref{sec:spectral_DSE} and \Cref{app:regularization}, and in particular  \labelcref{eq:lambda-aQuantum,eq:Quantumnu}. 

In \labelcref{eq:microscopic-action}, we also introduced the chemical potential of the fermion species, $\mu_{\sigma}$ with $\sigma = (\uparrow, \downarrow)$. The fermionic fields $\psi_{\sigma}(\tau,\boldsymbol{x})$ are Grassmann valued and depend on the Euclidean time $\tau$, which is restricted to the circumference $\beta=1/T$, and the spatial coordinates $\boldsymbol{x}$. In \labelcref{eq:microscopic-action}, we have also used natural units, $\hbar = k_B = 2\,m_\psi = 1$, where $m_\psi$ is the mass of fermions. 

The emergence of bosonic dimers and their condensation  at low temperatures in the fermionic ultracold gas can be efficiently accommodated by rewriting the action \labelcref{eq:microscopic-action} with a Hubbard–Stratonovich transformation. Thus, we will consider a spin-balanced system ($\mu_{\uparrow}=\mu_{\downarrow}=\mu$), where the contact interaction of fermions with opposite spin is substituted by the exchange of bosonic dimers, 
\begin{align}
    S[\psi, \phi] = \int_{0}^{\beta} d\tau &\int d^3 x\, \Big[ \sum_{\sigma=\uparrow,\downarrow}
    \psi^*_{\sigma} (\partial_{\tau} - \nabla^2 - \mu) \psi_{\sigma} \notag \\
    &+ \nu \phi^* \phi - h \left(\phi^* \psi_{\uparrow} \psi_{\downarrow} - \phi \psi^*_{\uparrow} \psi^*_{\downarrow}\right) \,
    \Big] \,,
 \label{eq:BosonisedAction}
 \end{align}
where $h$ is the Feshbach coupling between the fermions and bosons, and $\nu$ is the detuning of the dimer. On the mean field level, similarly to \labelcref{eq:lambda-a}, the couplings and the detuning parameter would be related by  
\begin{align} 
	\lambda = - \frac{h^2}{\nu} \,. 
\label{eq:HSrelation}
\end{align}
For the full relation on the quantum level, see \Cref{sec:spectral_DSE} and \Cref{app:regularization}, and in particular \labelcref{eq:lambda-aQuantum,eq:Quantumnu}.  

It is readily shown that \labelcref{eq:BosonisedAction} reduces to \labelcref{eq:microscopic-action} on the equation of motion (EoM) of $\phi$, also using \labelcref{eq:HSrelation}. Solving \labelcref{eq:HSrelation} for the detuning parameter on the full quantum level, leads to \labelcref{eq:Quantumnu} in \Cref{app:regularization}. Notably, while the four-fermion coupling $\lambda$ gets strongly affected by fluctuations, the Feshbach coupling $h$ is well-approximated by its classical value, as these fluctuation effects are encoded in $\nu$. The latter is used to tune the (inverse) scattering length, for a detailed discussion of all these properties see~\cite{Diehl:2009ma}. We also emphasise that the absence of an explicit $s$-channel four-fermion scattering can be sustained on the quantum level within the fRG approach with emergent composites or rebosonisation, for the respective conceptual developments see \cite{Gies:2001nw, Pawlowski:2005xe, Floerchinger:2009uf, Fu:2019hdw}, for an application to the present system see \cite{Floerchinger:2008qc, Floerchinger:2009pg}. Heuristically, this approach may be understood as a scale-dependent Hubbard-Stratonovich transformation. 

For a dilute ultracold Fermi gas, the effective range $r_0$ of the interaction is of the order of the inverse van der Waals length $l_{\mathrm{vdw}}$ and defines a physical momentum cutoff scale $\Lambda\sim 1/r_0$, which is much larger then all relevant physical scales ($\Lambda\gg k_F$). Throughout this work, we will work in the limit of contact interactions with zero effective range, such that $\Lambda$ can be sent to infinity in the end. For a more detailed discussion of the renormalisation scheme, we refer to~\Cref{app:regularization}. 

The complex bosonic field $\phi(\tau,\bm{x})$ has been introduced as an efficient book-keeping device for the $s$-channel interaction between fermions of opposite spin. For example, bubble resummations of $s$-channel diagrams, but also beyond bubble resummations of $t$ and $u$-channel contributions are encoded in two-point function diagrams or corrections to the Yukawa coupling. More importantly, emergent dynamical $s$-channel dimer degrees of freedom are already taken care of as is their condensation. 

Hence, by tuning the dimensionless interaction strength $(k_Fa)^{-1}$, the system can be driven from a BCS-type superfluid to a Bose-Einstein condensed state, allowing for the exploration of the BCS-BEC crossover~\cite{Nozieres:1985zz,SadeMelo:1993zz,Chen2005,PhysRevB.66.024510}. Here, the interaction strength is measured in terms of the Fermi momentum 
\begin{align}
k_F=(3\pi^2n)^{1/3}\,,  \qquad T_F=\varepsilon_F=k_F^2\,, 
\label{eq:kFTF}
\end{align}
where $n$ is the total density, and we have also defined the Fermi energy/temperature for later use. On the BCS side of the crossover, $(k_Fa)^{-1}<0$, the bosonic dimer $\phi$ describes weakly bound Cooper pairs. On the BEC side, $(k_Fa)^{-1}>0$, $\phi$ describes tightly bound bosonic molecules. Bose-Einstein condensation of the bosonic pairs, \ie, a non-vanishing expectation value of $\phi$, leads to superfluidity in the system. In the following, we will consider the normal phase of the BCS-BEC crossover with a special focus on the strongly correlated, unitary regime at $(k_Fa)^{-1}=0$.

\section{Spectral Dyson-Schwinger Equations} 
\label{sec:spectral_DSE}

In this Section we briefly introduce the coupled spectral propagator DSEs which will be used in obtaining all numerical and analytical results in this work. We also specify the underlying approximation of the effective action.

\subsection{Gap equations and the approximation of the effective action}
\label{sec:Gap+EffAct}

Functional approaches, such as the DSEs, the fRG or two-particle irreducible (2PI) resummation schemes, provide exact non-perturbative relations for correlation functions in terms of full propagators and vertices (fRG) as well as the classical vertices (DSE, 2PI). For reviews, see~\cite{Swanson:2010pw} (DSEs) and~\cite{Kopietz:2010zz, Metzner:2011cw, Scherer:2010sv, Boettcher:2012cm, Dupuis:2020fhh} (fRG) and~\cite{Berges:2004yj} (2PI). The central object in these approaches is the quantum effective action $\Gamma[\Phi]$, which is the generating functional of one-particle irreducible (1PI) correlation functions, or the 2PI effective action. Correlation functions of $n$ fields are obtained by $n$ functional derivatives with respect to the the mean fields, $\Phi=(\psi_{\sigma},\psi_{\sigma}^*,\phi,\phi^*)$, where for the sake of notational simplicity we have used the same expressions as for the classical fields. 

\begin{figure}[t]
	\centering
	\includegraphics[width=\linewidth]{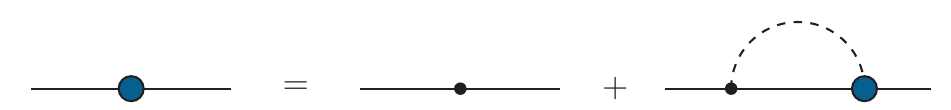}\vspace*{0.5cm}
	\includegraphics[width=\linewidth]{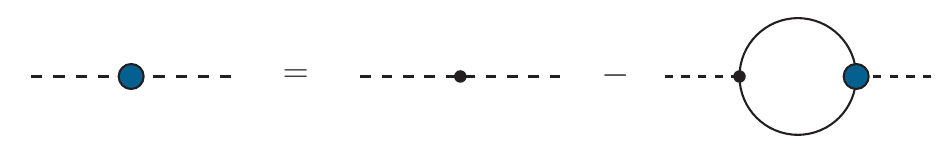}
	\caption{Fermion (solid line) and boson (dotted line) propagator Dyson-Schwinger equation for the balanced ultracold Fermi gas. Notation as defined in~\Cref{fig:dse_notation}.}
	\label{fig:DSE}
\end{figure}

\begin{figure}[t]
	\centering
	\includegraphics[width=\linewidth]{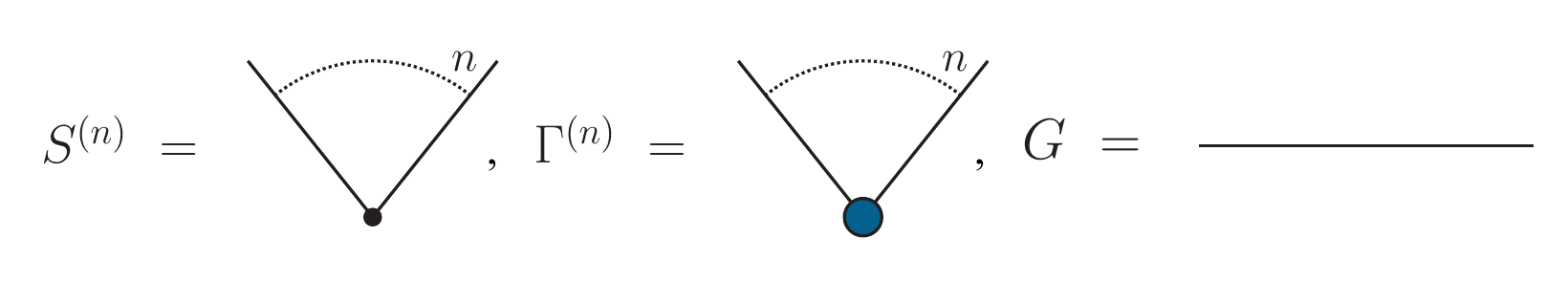}
	\caption{Diagrammatic notation used throughout this work: Lines stand for full propagators, small black dots stand for classical vertices, and larger blue dots stand for full vertices.}
	\label{fig:dse_notation}
\end{figure}

The most important example is the full non-perturbative two-point function, the full propagator $G_{\Phi}$, which is the pivotal building block of all diagrammatic functional approaches. In terms of the 1PI effective action, it is given  is the inverse of the 1PI two-point function, $G_{\Phi} = \left[\Gamma^{(2)}\right]^{-1}_{\Phi\Phi^*}$, which is a matrix inverse in field space. In the normal phase, the expectation value of the bosonic field is zero, \ie $\phi=0$, and the propagator, if evaluated on the EoMs, is diagonal in field space. Since we are dealing with a balanced system, the propagators of the $\uparrow$ and $\downarrow$ species are the same and we are left with only one fermion propagator $G_{\psi_{\uparrow}}=G_{\psi_{\downarrow}}=G_{\psi}$ and one boson propagator $G_{\phi}$. The gap equation or DSE for the inverse propagator $\Gamma^{(2)}_{\Phi\Phi^*}$ involves the classical inverse propagator $S^{(2)}_{\Phi\Phi^*}$, whose fermionic and bosonic components are obtained from the microscopic action~\labelcref{eq:BosonisedAction}, 
\begin{align} 
	S^{(2)}_{\psi\psi^*}(P) = -i\omega_n+\bm{p}^2-\mu \,, \qquad
	S^{(2)}_{\phi\phi^*}(Q) = \nu \,. 
\label{eq:classical-propagators}
\end{align}
Here, $P=(\omega_n,\bm{p})$ and $Q=(\epsilon_n,\bm{q})$ with the fermionic ($\omega_n$) and bosonic ($\epsilon_n$) Matsubara frequencies respectively, 
\begin{align}
\omega_n=(2n+1)\pi T\,,\qquad \epsilon_n=2n\pi T\,. 
\label{eq:FermBosMatsubara}
\end{align}
The coupled Dyson-Schwinger equations for the fermion and boson propagator are depicted in~\Cref{fig:DSE}, while the diagrammatic notation is summarised in~\Cref{fig:dse_notation}. The renormalised fermion and boson gap equations are given by
\begin{align}
	G^{-1}_{\psi}(P) &= -i\omega_n+\bm{p}^2-\mu + \Sigma_{\psi}(P) \,, \notag \\[1ex]
	G^{-1}_{\phi}(Q) &= -\frac{h^2}{8\pi a} - \Pi_{\phi}(Q) \,,
 \label{eq:coupled_DSEs}
 \end{align}
with the renormalised fermionic and bosonic self-energy, $\Sigma_{\psi}(P)$ and $\Pi_{\phi}(Q)$, respectively. For more details on the spectral renormalisation procedure see~\Cref{app:regularization}.

Now we specify the approximation of the effective action used for the following computations. As already discussed below \labelcref{eq:HSrelation}, the approximation of a classical constant Yukawa coupling $h$ is working well as the fluctuations of the four-fermi interaction are well-captured by that in the detuning parameter. The latter is part of the full propagator and hence will be accessed here. This approximation is readily lifted by also solving the spectral DSE for the Yukawa coupling and the remnant four-fermion interaction which will be considered elsewhere. This leads us to the following approximation of the renormalised finite effective action, 
\begin{align}\nonumber 
	\Gamma[\psi, \phi] = &\,\int_{0}^{\beta} d\tau \int d^3 x\, \Biggl[ 
	\psi^*_{\sigma} \Bigl(\partial_{\tau} - \nabla^2 - \mu + \Sigma_\psi\Bigr) \psi_{\sigma} 	\\[1ex] 
	 &\hspace{-.3cm}+\phi^*  \Bigl(
	 \nu - \Pi_\phi\Bigr)\phi - h  \Bigl(\phi^* \psi_{\uparrow} \psi_{\downarrow} - \phi \psi^*_{\uparrow} \psi^*_{\downarrow}\Bigr) \,
	\Biggr] \,,
	\label{eq:EffAct}
\end{align}
where the spin sum over $\sigma=\uparrow,\downarrow$ is implied. \Cref{eq:EffAct} includes the full frequency and momentum dependent fermionic and bosonic propagator and keep the classical interaction part. This approximation incorporates the $s$-channel bubble-resummation with full fermionic propagators in $\Sigma_\psi$ as the bosonic dimer propagator simply constitutes the bubble chain. 

Using \labelcref{eq:BosonisedAction,eq:EffAct} on both sides of the gap equations \labelcref{eq:coupled_DSEs} leads us to the diagrammatic expression for the non-perturbative self-consistent and renormalised self-energies at finite temperature, 
\begin{align} \nonumber 
	\Pi_{\phi}(Q) = &\,h^2 \int_{\bm{p}} \left[ T\sum_{\omega_m} G_{\psi}(P) G_{\psi}(Q-P) - \frac{1}{2\bm{p}^2} \right] \,, \notag \\[1ex]
	\Sigma_{\psi}(P) = &\,h^2 \int_{\bm{q}}\, T\sum_{\epsilon_m} G_{\phi}(Q) G_{\psi}(Q-P) \,. 
\label{eq:self-energies}
\end{align}
with 
\begin{align}
	\int_{\bm{q}}=\int \frac{d^3q}{(2\pi)^3}\,, 
	\label{eq:Intd3p}
\end{align}
and similarly for the integral over $\boldsymbol{p}$. The subtraction with $-1/(2 \boldsymbol{p}^2)$ on the right hand side of the equation for $\Pi_\phi$ in  \labelcref{eq:self-energies} arises from the spectral renormalisation. \Cref{eq:self-energies} is given by \Cref{fig:DSE} with two classical vertices in each diagram. Note also, that in the fermionic self-energy $\Sigma_{\psi}(P)$, the sum is over bosonic Matsubara frequencies $\epsilon_m=2m\pi T$, whereas in the bosonic self-energy $\Pi_{\phi}(Q)$, the sum is over fermionic Matsubara frequencies $\omega_m=(2m+1)\pi T$.

Evidently, the present truncation with $\Gamma^{(3)} = S^{(3)}$ in the gap equations, \labelcref{eq:self-energies} derived from \labelcref{eq:EffAct}, is equivalent to an $s$-channel bubble-resummation, also known as the self-consistent T-matrix approximation~\cite{Hanai2014, PhysRevA.80.063612, PhysRevB.99.094502}. However, the present functional approach can be generalised straightforwardly in a self-consistent way by taking the full three-point function $\Gamma^{(3)}$ into account, as has been done in~\cite{Horak:2020eng,Horak:2023hkp} for a scalar theory. Respective results will be presented elsewhere.

\subsection{Spectral representation} \label{sec:spectral-representation}

The spectral representation of the propagator is at the core of the spectral functional approach developed in \cite{Horak:2020eng}. In the present work, we employ the Källén-Lehmann (KL) representation of the full propagators~\cite{PhysRevA.80.063612,Abrikosov1975,Fetter1971}
\begin{align} 
	G(\omega_n,\bm{p}) = \int_{-\infty}^{\infty} d\lambda \, \frac{\rho(\lambda,\bm{p})}{-i\omega_n+\lambda} \,,
\label{eq:spectral-representation}
\end{align}
where $\rho(\lambda,\bm{p})$ is the frequency and momentum-dependent spectral function. In this way, the spectral function acts as a linear response function of the propagator, encoding the energy spectrum of the theory. \Cref{eq:spectral-representation} leads to the following inverse relation between the spectral function and the retarded propagator,
\begin{align} 
	\rho(\omega,\bm{p}) = \frac{1}{\pi}\, \mathrm{Im}\, G^R(\omega,\bm{p}) \,,
\label{eq:spectral-relation}
\end{align}
where $G^R(\omega,\bm{p})=G(-i(\omega+i0^+),\bm{p})$ and $\omega$ is now a real frequency. Note that 
\labelcref{eq:spectral-representation} always defines the spectral function, related to the statistical function by the fluctuation-dissipation theorem, whereas \labelcref{eq:spectral-representation} only holds true for fields that define physical asymptotic states. While this is true in the present ultracold gas, it is far from obvious for theories defined in terms of unphysical degrees of freedom such as gauge fields, for more details see~\cite{Horak:2022aza, Horak:2022myj, Horak:2021pfr}. In summary, the present ultracold gas is optimally suited for application of spectral functional approaches. 

The existence of a spectral representation restricts all non-analyticities of the propagator to lie on the real frequency and momentum axis. The fermion spectral function satisfies the sum rule
\begin{align} \label{eq:spectral-sum-rule}
	\int_{-\infty}^{\infty} d\lambda\,\rho_{\psi}(\lambda,\bm{p}) = 1 \,.
\end{align}
The fermionic spectral functions satisfy $\rho_{\psi}(\omega,\bm{p}) \geq 0$. In our approach, the frequency and momentum dependence of the bosonic inverse propagator is only given by the self-energy, see \Cref{eq:coupled_DSEs}. Accordingly, its spectral function is not that of a physical state and is not normalised to unity as the fermionic one. Moreover, the respective spectral function is not positive semi-definite, but satisfies $\mathrm{sgn}(\omega) \rho_{\phi}(\omega,\bm{p})\geq 0$~\cite{Fetter1971}. Note that the negative sign of the boson spectral function for negative frequencies guarantees the positivity of the boson momentum distribution function~\cite{PhysRevA.88.013627}. 

Using the spectral representation~\labelcref{eq:spectral-representation} for the fermion and boson propagator, the self-energies can be written in terms of spectral loop integrals. We find for the bosonic self-energy 
\begin{align} \nonumber 
	\Pi_{\phi}(\epsilon_n,\bm{q}) = h^2\int_{\bm{p}}\left[\int_{\lambda_1,\lambda_2} \rho_{\psi}(\lambda_1,\bm{p}) \rho_{\psi}(\lambda_2,\bm{q}-\bm{p}) \right.\\[1ex]
	\left. \times\, T\sum_{\omega_m} \frac{1}{i\omega_m-\lambda_1}\frac{1}{i(\epsilon_n-\omega_m)-\lambda_2} - \frac{1}{2\bm{p}^2} \right] \,. 
	\label{eq:spectral-boson-self-energy}
\end{align}
The fermionic self-energy takes the form  
\begin{align} 
	\Sigma_{\psi}(\omega_n,\bm{p}) =&\, h^2\int_{\lambda_1,\lambda_2,\bm{q}} \rho_{\phi}(\lambda_1,\bm{q}) \rho_{\psi}(\lambda_2,\bm{q}-\bm{p}) \notag \\[1ex]
	&\times T\sum_{\epsilon_m} \frac{1}{i\epsilon_m-\lambda_1}\frac{1}{i(\epsilon_m-\omega_n)-\lambda_2} \,.  
\label{eq:spectral-fermion-self-energy}
\end{align}
In both, \labelcref{eq:spectral-boson-self-energy,eq:spectral-fermion-self-energy}, we have used \labelcref{eq:Intd3p} and 
\begin{align}
\int_{\lambda}=\int_{-\infty}^{\infty}d\lambda\,. 
\label{eq:Intlambda}
\end{align} 
Importantly, the Matsubara sum in~\labelcref{eq:spectral-fermion-self-energy} and~\labelcref{eq:spectral-boson-self-energy} can be carried out analytically. This leaves us with symbolic expressions in terms of the argument $\epsilon_n$ and $\omega_n$ for both self-energies that can be evaluated at any complex frequency. The explicit spectral integral expressions for $\Sigma_{\psi}$ and $\Pi_{\phi}$ can be found in \Cref{app:boson-self-energy-calculation} and~\labelcref{app:fermion-self-energy-calculation}.

\subsection{Evaluation at real frequencies} \label{sec:evaluation-real-frequencies}

The regularised and coupled DSEs in~\Cref{eq:coupled_DSEs} can be evaluated for arbitrary complex frequencies. For the extraction of the spectral functions with~\labelcref{eq:spectral-relation}, we choose $\omega_n=-i(\omega+i\varepsilon)$ with $\varepsilon\rightarrow 0^+$. The limit $\varepsilon\rightarrow 0^+$ is performed analytically using the relation
\begin{align} \label{eq:principal-value}
	\frac{1}{x\pm i0^+} = P\left(\frac{1}{x}\right) \mp i\pi\delta(x) \,,
\end{align}
where $P(1/x)$ denotes the principal value of $1/x$. This allows us to write the imaginary part of the retarded self-energies as
\begin{align} \label{eq:imaginary-part-boson-self-energy}
	\mathrm{Im}\,\Pi^R_{\phi}(\omega,\bm{q}) &= \pi h^2\int_{\lambda,\bm{p}} \rho_{\psi}(\omega-\lambda,\bm{p}) \rho_{\psi}(\lambda,\bm{q}-\bm{p}) \notag \\[1ex]
	&\times \left[1-n_F(\omega-\lambda)-n_F(\lambda)\right] \,,
\end{align}
\begin{align} \label{eq:imaginary-part-fermion-self-energy}
	\mathrm{Im}\,\Sigma^R_{\psi}(\omega,\bm{p}) &= \pi h^2\int_{\lambda,\bm{q}} \rho_{\phi}(\omega+\lambda,\bm{q}) \rho_{\psi}(\lambda,\bm{q}-\bm{p}) \notag \\[1ex]
	&\times \left[-n_B(\omega+\lambda)-n_F(\lambda)\right] \,. 
\end{align}
Here we have introduced the Fermi-Dirac distribution $n_F$ and Bose-Einstein distribution $n_B$ with 
\begin{align}
n_F(x)=\frac{1}{e^{x/T}+1} \,,\qquad n_B(x)=\frac{1}{e^{x/T}-1}\,.
\label{eq:nFnB}
\end{align}
A detailed derivation of the above formulae is deferred to~\Cref{app:boson-self-energy-calculation} and~\Cref{app:fermion-self-energy-calculation}. Note that the pole of $n_B(\omega+\lambda)$ in the fermion self-energy is exactly cancelled by the zero crossing of the boson spectral function $\rho_{\phi}$, see~\Cref{sec:spectral-representation}.

From the imaginary part of the fermion self-energy, \eg, the real part can be obtained via the Kramers-Kronig relation
\begin{align} \label{eq:kramers-kronig}
	\mathrm{Re}\,\Sigma^R_{\psi}(\omega,\bm{p}) &= \frac{1}{\pi}\, P\int_{\lambda} \frac{\mathrm{Im}\, \Sigma^R_{\psi}(\lambda,\bm{p})}{\lambda-\omega} \,.
\end{align}
For the real part of the boson self-energy, the renormalisation of $\nu$ has to be taken into account. The problematic vacuum part can be treated analytically, see \Cref{app:implementation}. This set of equations allow for an iterative calculation of the retarded self-energies and therefore, a direct determination of the spectral functions which are presented in the next section. For a detailed discussion of the numerical implementation, see \Cref{app:implementation}.

\section{Results} \label{sec:results}

In this Section, we discuss the numerical results for the fully self-consistent spectral functions as well as its use for the computation of the density equation of state, the Tan contact and ejection rf spectra at unitarity. In \Cref{sec:spectral-functions} we focus on the novel results obtained in real frequencies. In \Cref{sec:rf-spectra} we use the spectral functions from \Cref{sec:spectral-functions} for computing radio-frequency (rf) spectra in comparison to experimental data. Finally, in \Cref{sec:density_eos} we compute the equation of state and use the large momentum dependence of the density for extracting the Tan contact.

\subsection{Spectral functions}
 \label{sec:spectral-functions}

We present numerical results for the self-consistent spectral functions of the spin-balanced Fermi gas in the normal phase at different scattering lengths and, in particular, at unitarity. Physical properties of the single-particle fermion spectral function have been discussed extensively in the literature~\cite{PhysRevA.91.043627, PhysRevB.85.024517, PhysRevA.80.063612}. In this work, we also discuss the bosonic two-particle (dimer) spectral function which has not received much attention in the past. For details on the numerical implementation, we refer the reader to \Cref{app:implementation}.

\begin{figure}[t]
	\centering
	\includegraphics[width=0.98\linewidth]{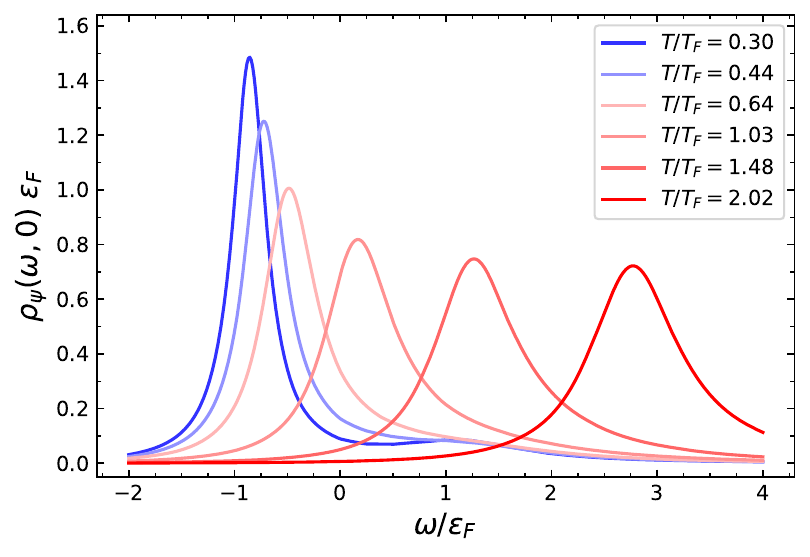}
	\caption{Results for the fermionic spectral function $\rho_{\psi}(\omega,\bm{0})\,\varepsilon_F$ for the unitary Fermi gas at different temperatures, which are used for the computation of the rf spectra.}
	\label{fig:fermion-specs-unitary}
\end{figure}

\begin{figure*}[t]
	\centering
	\subfloat[$(k_Fa)^{-1}=-0.5$]{\includegraphics[width=0.32\textwidth]{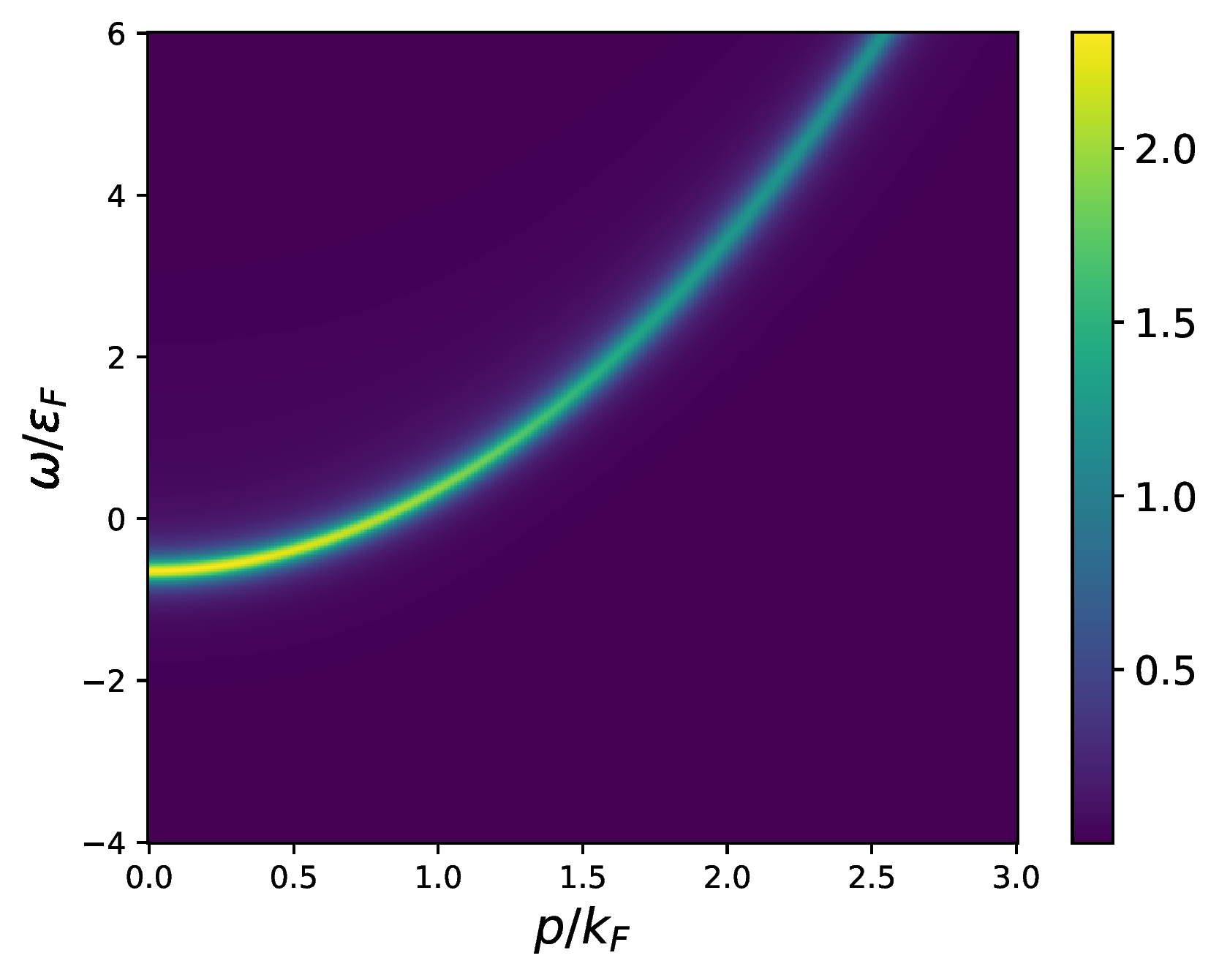}}
	\subfloat[$(k_Fa)^{-1}=0$]{\includegraphics[width=0.32\textwidth]{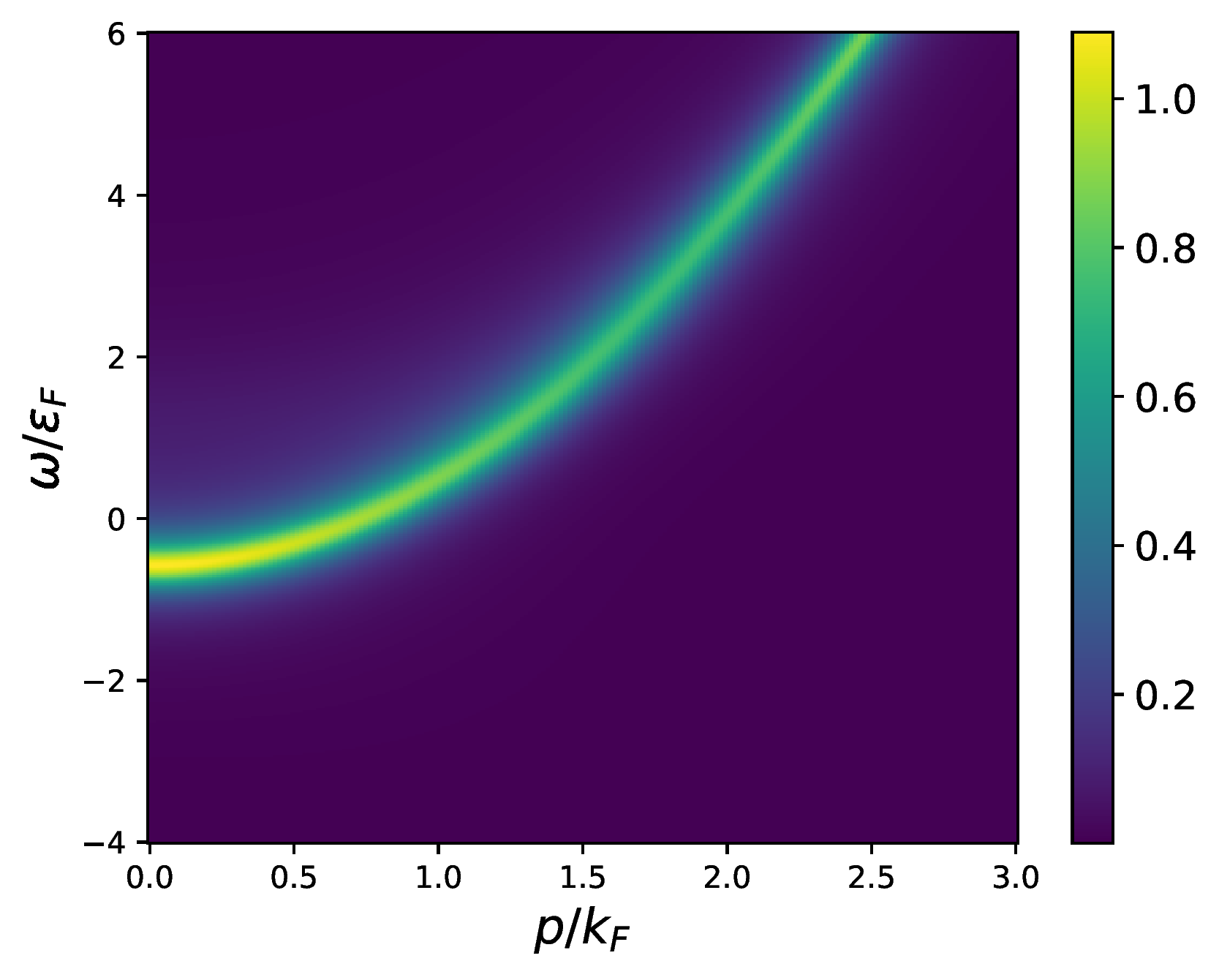}}
	\subfloat[$(k_Fa)^{-1}=0.5$]{\includegraphics[width=0.32\textwidth]{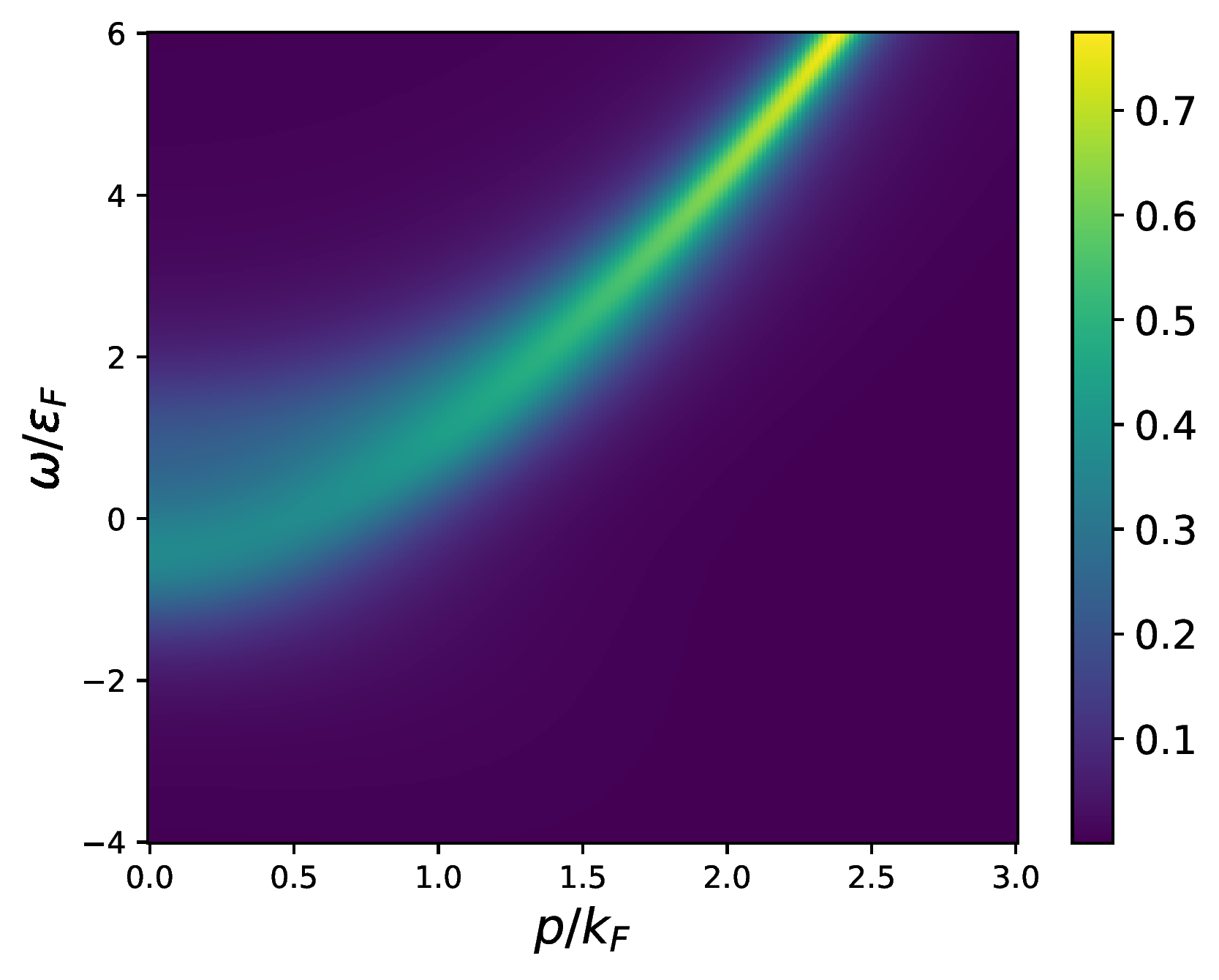}}
	\caption{Results for the fermionic spectral function $\rho_{\psi}(\omega,\bm{p})\,\varepsilon_F$ for $T/T_F=0.56$ at different interaction strengths $(k_Fa)^{-1}$. \\
		(a) BCS regime: $\beta\mu=0.5$ (b) unitarity: $\beta\mu=0.13$ (c) BEC regime: $\beta\mu=-0.54$.}
	\label{fig:fermion-specs}
\end{figure*}

\begin{figure*}[t]
	\centering
	\subfloat[$(k_Fa)^{-1}=-0.5$]{\includegraphics[width=0.325\textwidth]{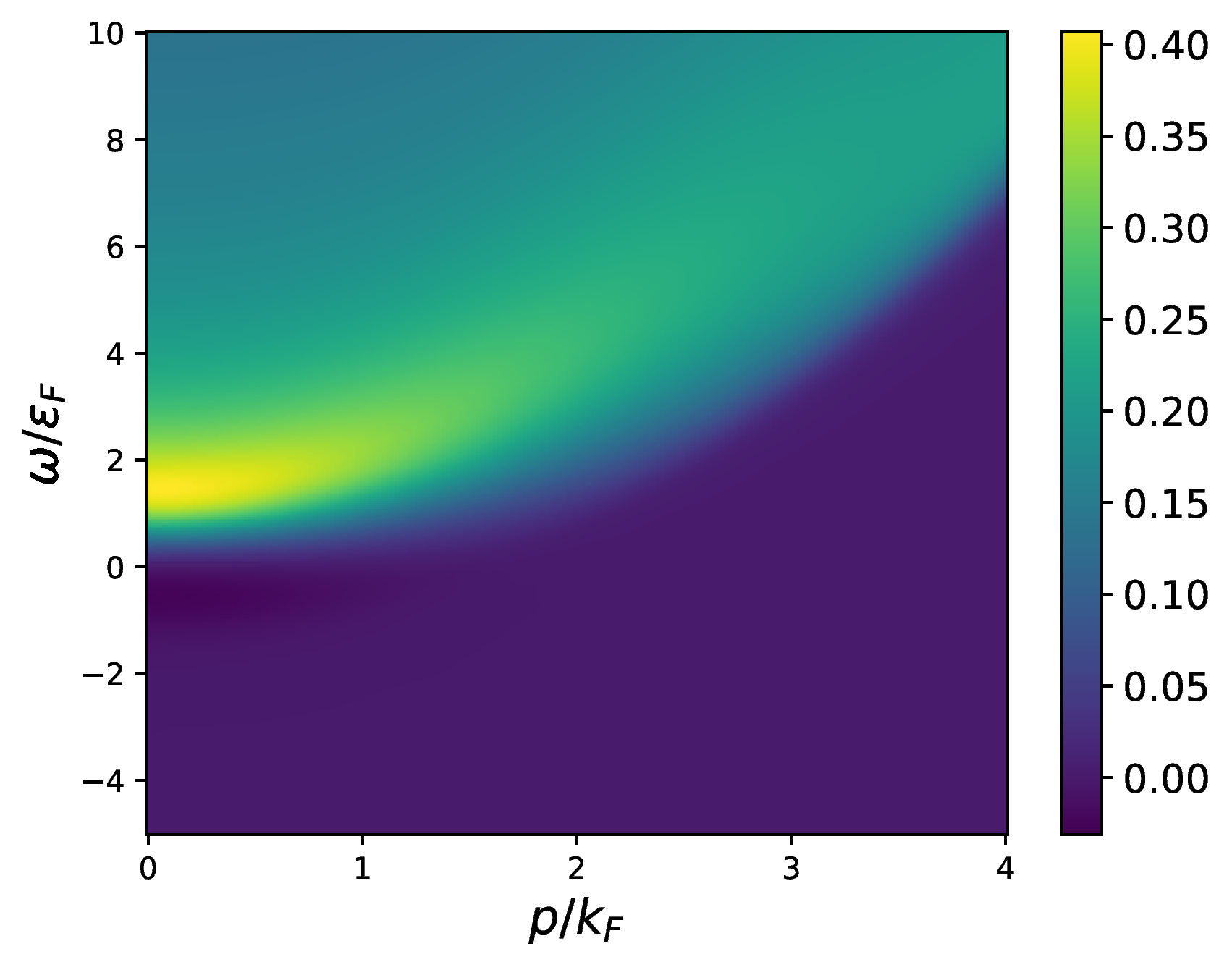}}
	\subfloat[$(k_Fa)^{-1}=0$]{\includegraphics[width=0.32\textwidth]{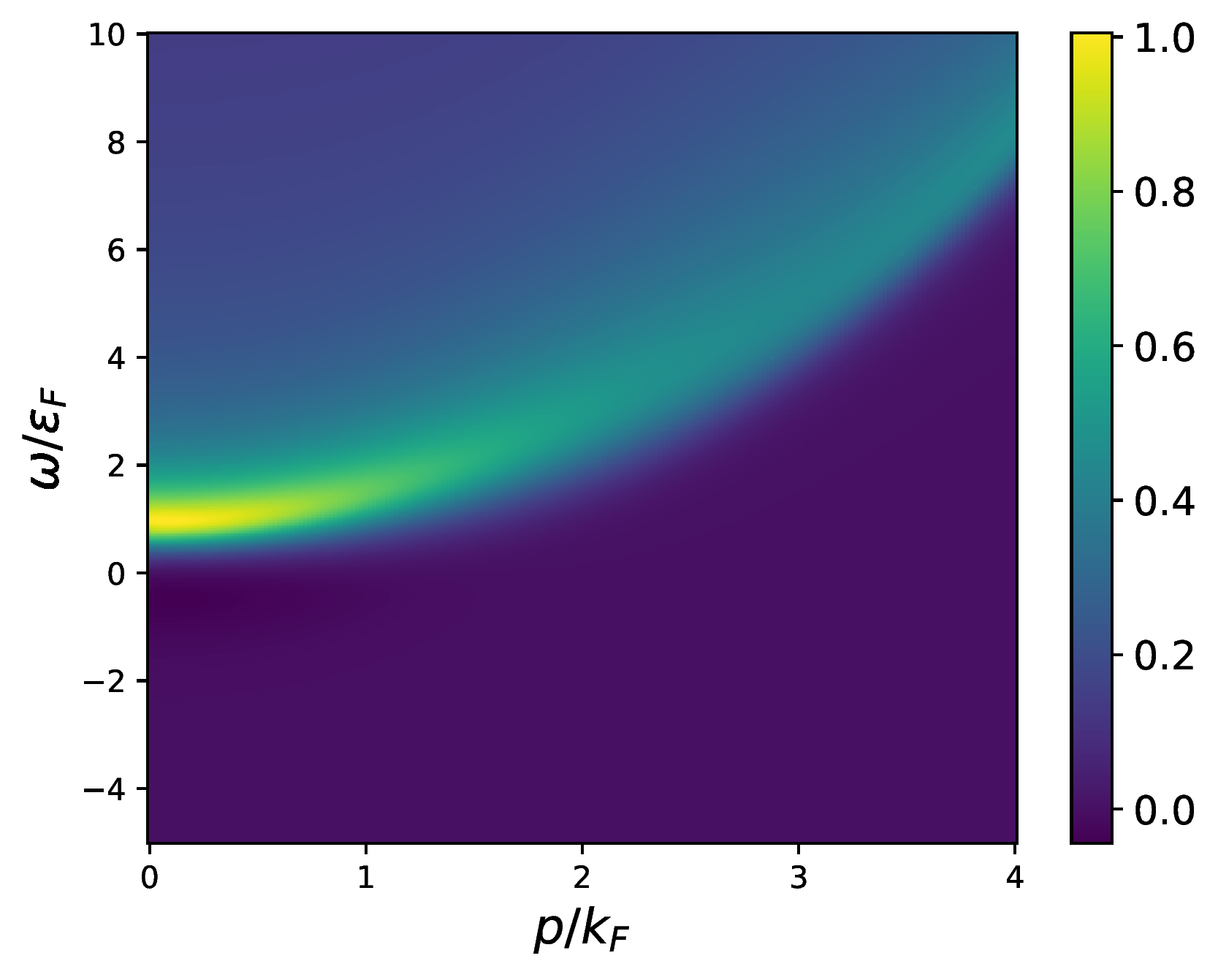}}
	\subfloat[$(k_Fa)^{-1}=0.5$]{\includegraphics[width=0.32\textwidth]{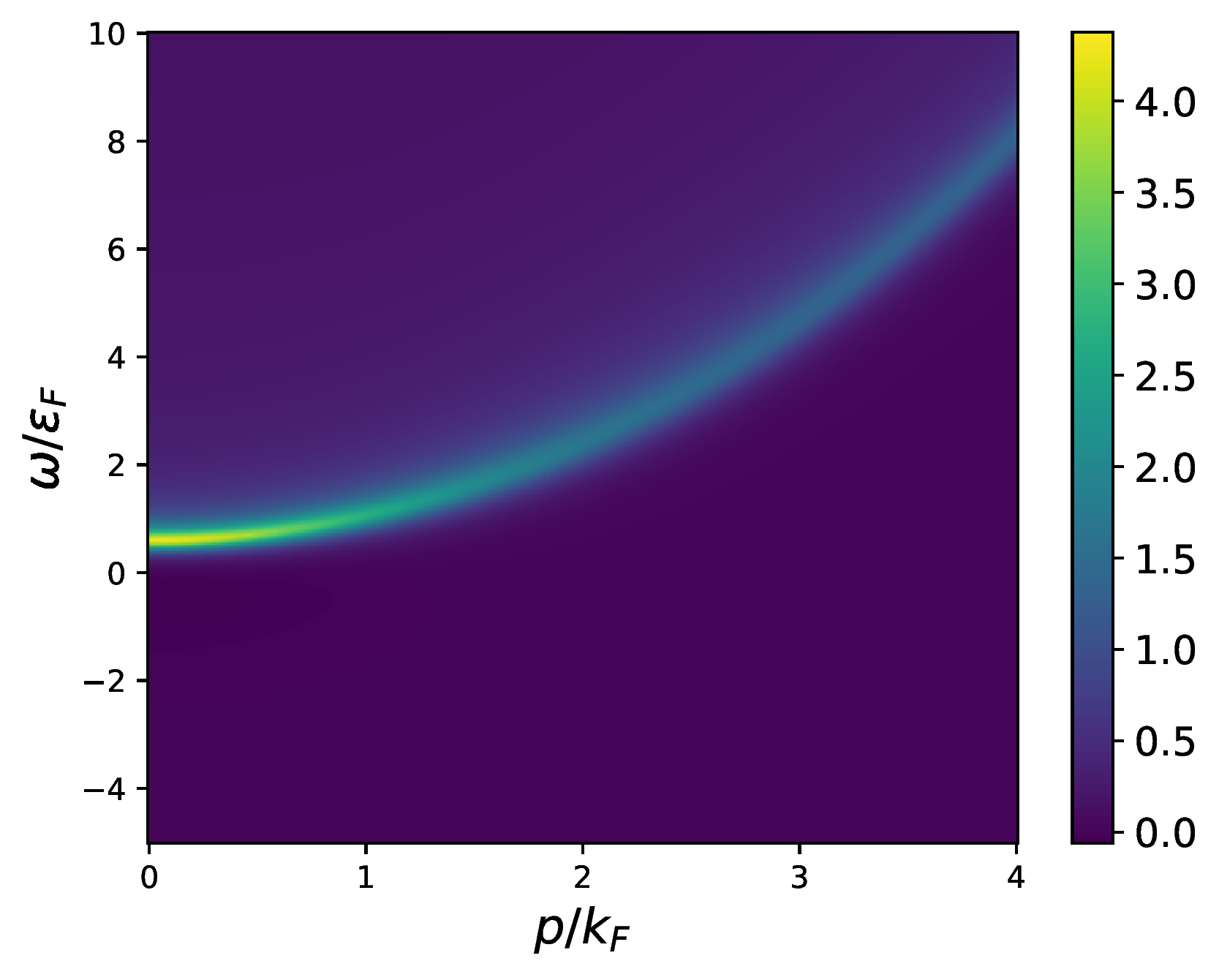}}
	\caption{Results for the bosonic dimer spectral function $h^2\rho_{\phi}(\omega,\bm{p})\,\sqrt{\varepsilon_F}/(8\pi)$ for $T/T_F=0.56$ at different $(k_Fa)^{-1}$. \\
		(a) BCS regime: $\beta\mu=0.5$ (b) unitarity: $\beta\mu=0.13$ (c) BEC regime: $\beta\mu=-0.54$.}
	\label{fig:boson-specs}
\end{figure*}

\Cref{fig:fermion-specs-unitary} shows results for the fermionic spectral function $\rho_{\psi}$ at $\bm{p}=0$ for the unitary Fermi gas at different temperatures, which are used for the computation of the rf spectra in \Cref{sec:rf-spectra}. Furthermore, \Cref{fig:fermion-specs} shows results for the full frequency and momentum-dependent fermion spectral function $\rho_{\psi}$, and \Cref{fig:boson-specs} shows the corresponding bosonic dimer spectral functions $\rho_{\phi}$ for different interaction strengths $1/(k_F a)$, with the Fermi momentum $k_F$ defined in \labelcref{eq:kFTF}. In \Cref{fig:fermion-specs-unitary,fig:fermion-specs,fig:boson-specs}, the frequency $\omega$ and momentum $\bm{p}$ are measured in the Fermi energy $\varepsilon_F$ and $k_F$, respectively, and the fermionic (bosonic) spectral functions have units of $\varepsilon_F^{-1}$ ($\varepsilon_F^{-1/2}$). We state all results in dimensionless form. Note that the bosonic dimer spectral function is not normalised and depends on the choice of the Feshbach coupling $h$. This is connected to it being a pure interaction exchange boson and not a real particle. To eliminate the $h$-dependence, we multiply $\rho_{\phi}$ by $h^2/(8\pi)$ since $h^2G_{\phi}/(8\pi)$ is the relevant quantity, which is related to the scattering amplitude, see~\Cref{app:regularization}.

Our results are computed directly in real-time and agree qualitatively with the reconstruction results from previous works~\cite{PhysRevA.80.063612} for the fermion spectral function. Thus, our results corroborate the use of reconstruction methods for spectral functions. It is well-known that numerical reconstruction problems are ill-conditioned or at least have a high condition number. Moreover, the respective ambiguities are specifically large in the low frequency regime at finite temperature. The computation of dynamical properties such as the shear viscosity requires a precise determination of the low frequency regime and its transport peaks, which can be achieved with our spectral functional approach. We envisage a two-step process with a direct use of Euclidean benchmark results for a quality check of the direct spectral computation as well as using its structural low frequency properties as an inductive bias for spectral reconstructions.  

The results for the bosonic dimer spectral function encode important information about the physics of the ultracold Fermi gas and are consistent with previous studies. First, we observe a very broad peak structure for weak attractive interactions in the BCS regime, and a very sharp peak structure with thermal broadening for strong attractive interactions in the BEC regime. The latter property signals the tightly bound molecules on the BEC side. On this side of the BCS-BEC phase diagram at $1/(k_Fa)=0.5$, the boson spectral function is very sharp and the system can be described as normal Bose liquid. On the BCS side of the crossover at $1/(k_Fa)=-0.5$, the fermion spectral function is sharper and the system is described by a normal Fermi liquid.

\begin{figure}[t]
	\centering
	\includegraphics[width=0.954\linewidth]{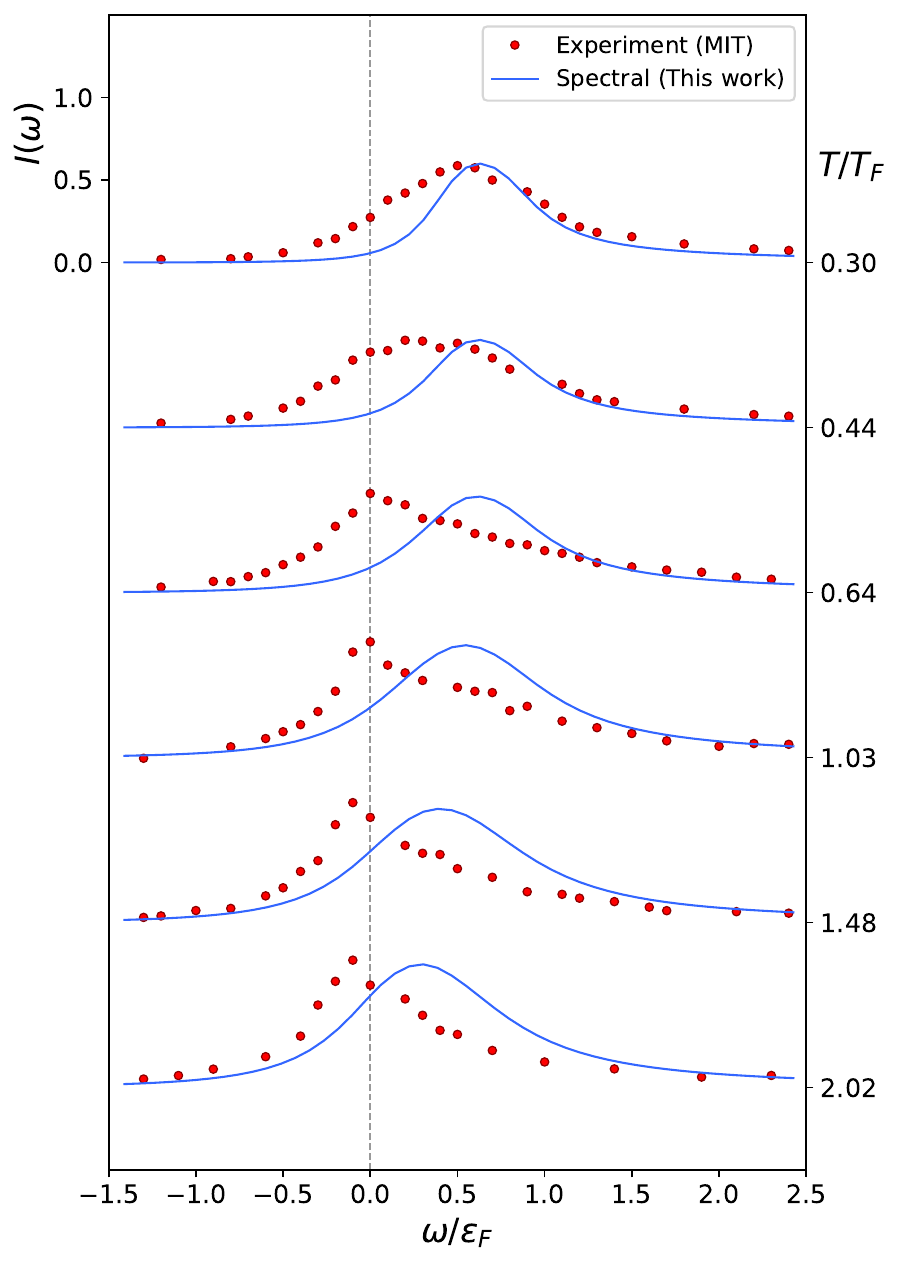}
	\caption{Calculated ejection rf spectra $I(\omega)$ for the spin-balanced unitary Fermi gas as a function of the reduced temperature $T/T_F$. Results of this work (solid lines) are compared to experimental data from MIT (circles)~\cite{PhysRevLett.122.203402}. A Fourier broadening of $0.1\varepsilon_F$ to account for the finite experimental resolution, and a right-shift by $0.09\varepsilon_F$ to account for the final state interaction were applied.}
	\label{fig:rf-spectra}
\end{figure}
Additionally, the boson spectral function reveals crucial information about the critical region of the phase transition to the superfluid state. The onset of superfluidity is marked by the divergence of the boson propagator at zero frequency and momentum, $G^{-1}_{\phi}(0,\bm{0})=0$. This property is known as the Thouless criterion~\cite{Thouless1960}. Thus, the closer and sharper the peak gets at zero frequency and momentum, the closer the system is to the phase transition, until the spectral function eventually diverges at the critical temperature. We refrain from a respective analysis and only present results in the normal phase. 

Computations at and below the phase transition are beyond the scope of the present paper and shall be considered elsewhere. Note, however, that respective work in the spectral functional approach has already been done in relativistic systems both with the DSE and the fRG \cite{Horak:2020eng, Horak:2023hkp, Eichmann:2023tjk}. The approximations there go beyond that used in the present work. In \cite{Horak:2020eng, Horak:2023hkp, Eichmann:2023tjk}, the full $s$-channel resummation of the four-point function has been considered, as well as additional diagrams that arise in the presence of a condensate $\phi_0$. In contrast to the Monte Carlo (MC) integration routine used here, we used efficient integration routines in \cite{Horak:2020eng, Horak:2023hkp, Eichmann:2023tjk}, aiming at optimal convergence for low-dimensional integrals. The use of adaptive MC integrators in the present work has to be understood as a test case for higher-dimensional computations as will arise for non-trivial vertex approximations beyond mere sums of momentum channels. This may be relevant in the presence of competing order effects. 

In summary we emphasise, that contrary to respective statements in the very recent work \cite{Johansen2023}, the numerical implementation of the present spectral functional approach certainly is not less efficient than that used there. Indeed, one of the reasons for setting up the spectral functional approach was its numerical efficiency, which by now has been proven in abundance in \cite{Fehre:2021eob, Horak:2020eng, Horak:2023hkp, Eichmann:2023tjk, Horak:2022aza, Horak:2022myj, Horak:2021pfr, Horak:2023rzx}. These successful tests include its convergence in coupled sets of integral equations, as well as the evaluation of bound state properties in the broken phase, as well as through the phase transition from the symmetric to the broken phase. Specifically, self-consistent real-time computations in the broken phase have so far solely been done in the present approach and standard Bethe-Salpeter type computations. We add that, given the structural similarities of the approaches used in \cite{Johansen2023}, as well as \cite{Enss2023}, to the spectral functional approach, it should be possible to extend the former approaches towards the broken regime along the lines in \cite{Horak:2020eng, Horak:2023hkp, Eichmann:2023tjk}.

\subsection{Radio-frequency spectroscopy} 
\label{sec:rf-spectra}

\begin{figure}[t]
	\raggedleft
	\includegraphics[width=\linewidth]{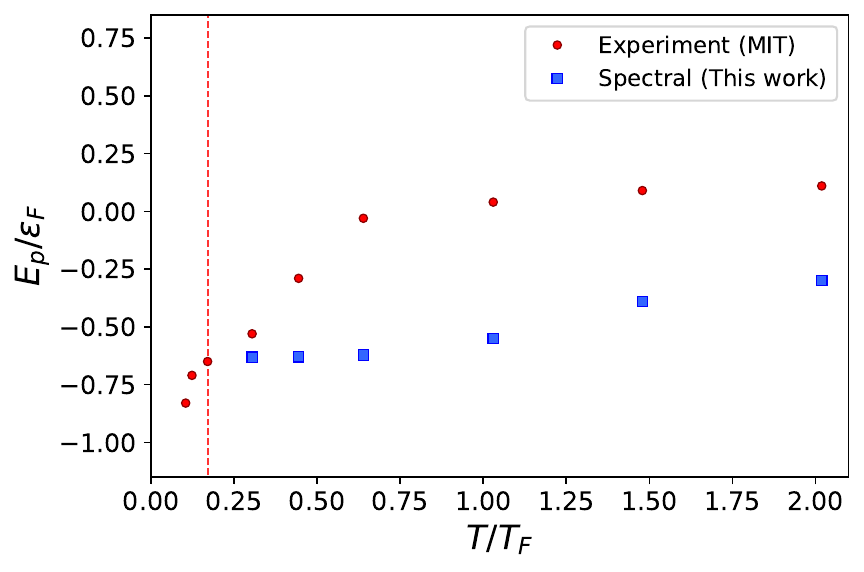}
	\includegraphics[width=0.975\linewidth]{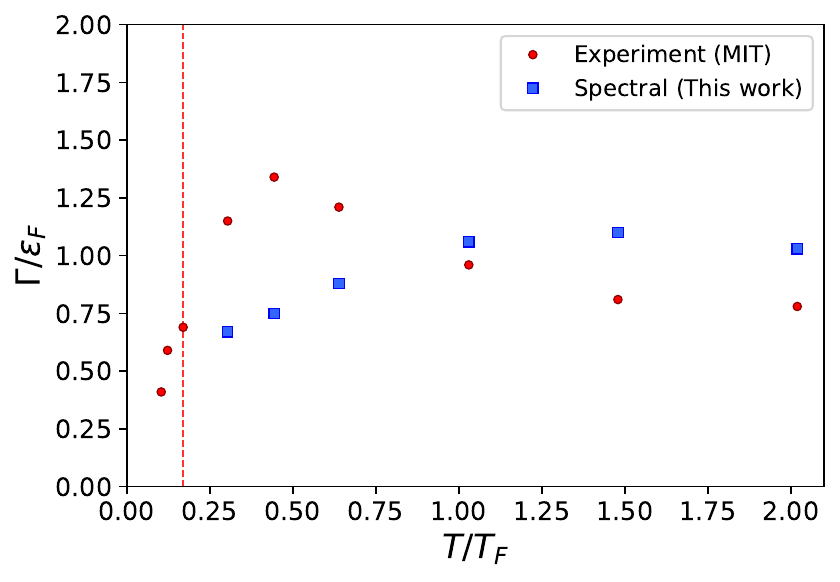}
	\caption{Peak positions ($E_p=-\omega_p$) and full width at half maximum $\Gamma$ as a function of the reduced temperature $T/T_F$. The red vertical dashed lines mark the superfluid phase transition.}
	\label{fig:rf-properties}
\end{figure}
Now we use our numerical results for the spectral functions to compute experimentally measurable radio-frequency (rf) spectra~\cite{PhysRevLett.99.170404,stewart2008using,Schneider2009}. We compare the results from our approach, obtained for the spin-balanced Fermi gas at unitarity with recent experimental data from MIT~\cite{PhysRevLett.122.203402}. Specifically, we apply the relation 
\begin{align} 
	I(\omega) = \int_{\bm{q}} \rho_{\psi}(\varepsilon_{\bm{q}}-\omega-\mu,\bm{q})\, n_F(\varepsilon_{\bm{q}}-\omega-\mu) \,,
\label{eq:rf-spectra}
\end{align}
for the computation of rf spectra $I(\omega)$ from the fermion spectral functions $\rho_{\psi}$, see ~\cite{PhysRevLett.99.170404, PhysRevA.82.033629}. In \Cref{eq:rf-spectra}, $\varepsilon_{\bm{q}}=\bm{q}^2$ is the classical fermion dispersion. 

Note that the chemical potential $\mu(T)$ is temperature-dependent and has to be determined self-consistently from the density equation~\labelcref{eq:density} below, see also~\cite{PhysRevA.78.033614,Haussmann:2007zz,PhysRevA.99.063606,Kagamihara:2019nsz,Pisani:2023bke}. More explicitly, the number density $n=1/(3\pi^2)$ is fixed by the choice $k_F=1$, and temperature is measured in units of $T_F$. Consequently, the chemical potential $\mu(T)$ has to be chosen such that the density remains constant. One readily observes that the rf spectrum is normalised to the total density $n$~\cite{PhysRevA.99.063606}, 
\begin{align} \label{eq:rf-dens}
	n = 2\int_{\lambda} I(\lambda) \,.
\end{align}
\Cref{fig:rf-spectra} shows our results in comparison with the experimental data from MIT~\cite{PhysRevLett.122.203402}. Apart from adjusting the peak heights, no fitting parameter have been used. In order to account for the finite rectangular rf pulse duration and, thus, a finite experimental resolution, a Fourier broadening of $0.1\varepsilon_F$ has been applied. Additionally, the curves have been right-shifted by an amount of $0.09\varepsilon_F$ to compensate residual final state interactions~\cite{PhysRevA.105.043303}. 

Even after taking into account all these possible factors, the calculated rf spectra do not fit the experimental data for higher temperatures very well. This is also apparent in the comparison of the peak position and full width as shown in \Cref{fig:rf-properties}. 

This discrepancy was already observed in~\cite{PhysRevA.105.043303} for the case of a highly spin-imbalanced unitary Fermi gas with the non-self-consistent T-matrix approach, and persists in the very recent update~\cite{Hu:2023}. It has been argued in~\cite{PhysRevA.105.043303}, the misalignment may be caused by the missing vertex corrections, that may also be relevant for pseudogap effects above $T_c$~\cite{Li:2023iqi}. Another possible explanation might be the absence of a trap average, see e.g.~\cite{PhysRevA.85.021602}.

\begin{figure}[t]
	\centering
	\includegraphics[width=\linewidth]{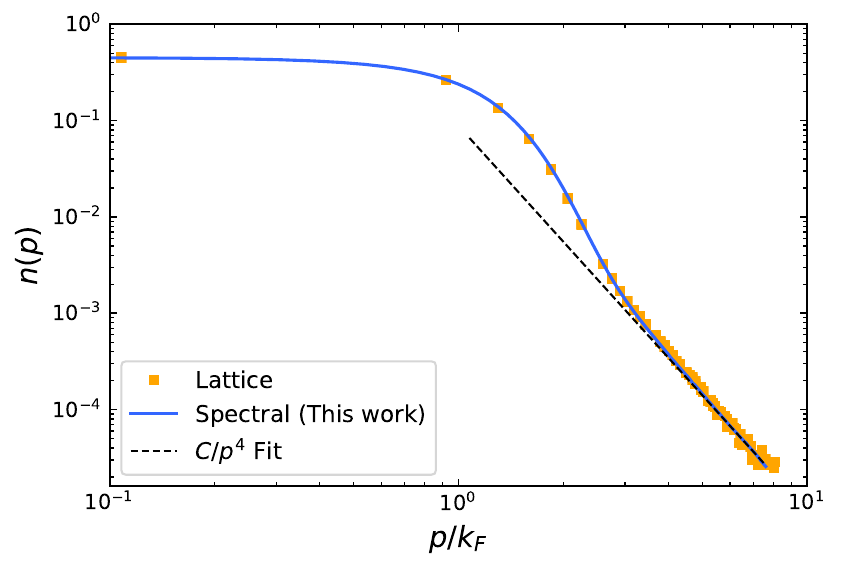}
	\caption{Large $p$ behaviour of the momentum density $n(p)$ of the balanced unitary Fermi gas at $T/T_F=0.9$ ($\beta\mu=-0.5$). The $1/p^4$ tail for $p\gg k_F$ is clearly visible and the contact $C$ can be determined from the fit. Lattice data from~\cite{MarcBauer}, see also~\cite{Drut:2010yn}.}
	\label{fig:momentum_density}
\end{figure}
%

\subsection{Equation of state and Tan contact}
\label{sec:density_eos}

Finally, we use the present results for the spectral functions to compute the equation of state for the unitary Fermi gas. Thermodynamic quantities, such as the total particle density, can be calculated precisely in imaginary frequencies without the need of analytic continuation. For this reason, it is a good way to validate the new spectral approach against well tested and robust imaginary-time calculations~\cite{Haussmann:2007zz,Frank2018,Drut:2010yn,Drut:2011tf,Drut:2012tg,Rammelmuller:2018hnk,Jensen:2019zkr}. 

\begin{figure}[t]
	\centering
	\includegraphics[width=\linewidth]{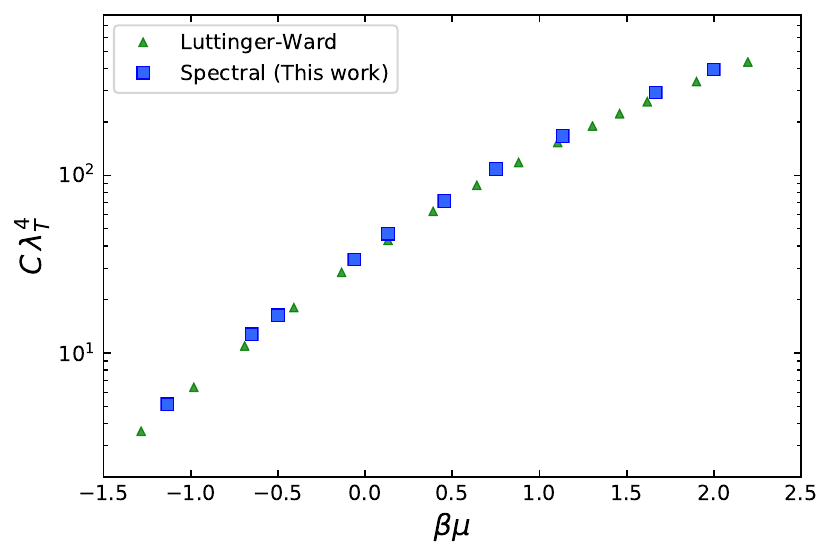}
	\caption{Universal contact $C\,\lambda^4_T$, where $\lambda_T=\sqrt{4\pi\beta}$ is the thermal wavelength, of the spin-balanced unitary Fermi gas as a function of dimensionless chemical potential $\beta\mu$. Our results in real frequencies compare well with Luttinger-Ward~\cite{Haussmann:2007zz,Frank2018} and Bold Diagrammatic Monte Carlo~\cite{Rossi:2013tik}. }
	\label{fig:contact_eos}
\end{figure}
\begin{figure}[t]
	\centering
	\includegraphics[width=\linewidth]{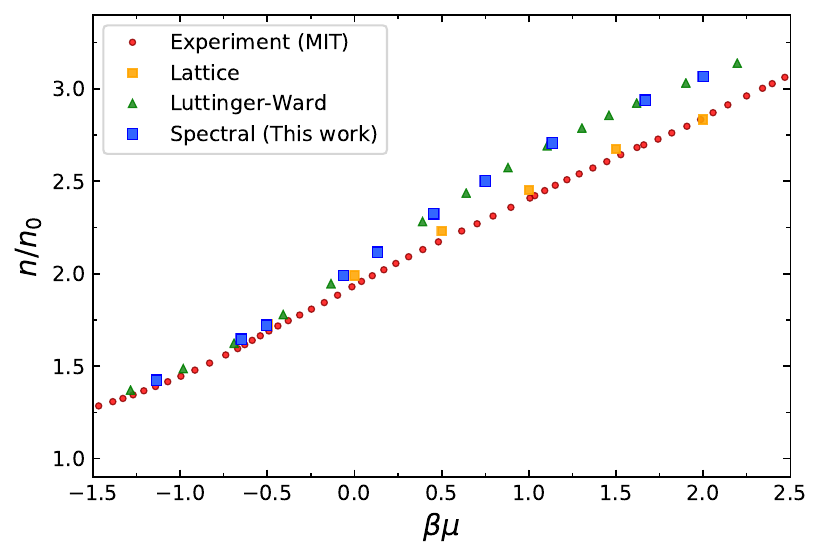}
	\caption{Normalised density $n/n_0$ of the spin-balanced unitary Fermi gas as a function of dimensionless chemical potential $\beta\mu$. Results directly obtained in real frequencies (this work) in comparison with experimental data from MIT~\cite{Ku2012}, lattice~\cite{MarcBauer}, see also \cite{Drut:2011tf,Rammelmuller:2018hnk}, and Luttinger-Ward~\cite{Haussmann:2007zz,Frank2018}.}
	\label{fig:density_eos}
\end{figure}

The total density $n$ of fermions at finite chemical potential $\mu$ and temperature $T$ can be calculated from the spectral function via \cite{Pisani:2023bke} 
\begin{align}
	n = 2\int_{\bm{p}}T\sum_{\omega_n}G_{\psi}(\omega_n,\bm{p})e^{i\omega_n0^+} = 2\int_{\bm{p}} n(\bm{p}) \,,
 \label{eq:density}
 \end{align}
where the factor two accounts for both fermion species and $n(\bm{p})$ is the momentum distribution function,
\begin{align} 
	n(\bm{p}) = \int_{\lambda}\rho_{\psi}(\lambda,\bm{p})\,n_F(\lambda) \,.
\label{eq:momentum-distribution}
\end{align}
For recent precision computations with functional approaches in the full phase structure of ultracold gases, see \cite{Faigle-Cedzich:2023rxd}. The large-momentum tail of $n(\bm{p})$ is related to the Tan contact~\cite{Tan2008,PhysRevLett.100.205301}
\begin{align}
	\lim_{p\rightarrow\infty} n(\bm{p}) = \frac{C}{p^4} \,,
 \label{eq:Tan-nas}
 \end{align}
where $p=\|\bm{p}\|$ and $C$ is the contact parameter that can be determined from the fit. For a different computation in the DSE approach, see \cite{Boettcher:2012dh}. We have compared the present results for the momentum density $n(\bm{p})$ and the total density $n$ with lattice data generated for this purpose with a Metropolis algorithm \cite{MarcBauer}. For results in the literature see also \eg~\cite{Drut:2010yn, Drut:2011tf,Rammelmuller:2018hnk}, using simulations with complex Langevin equations. The results for $n(\bm{p})$ in the present spectral DSE approach compare well to the benchmark results from the lattice simulation \cite{MarcBauer}. Moreover, in \Cref{fig:momentum_density} we also show a fit to the large momentum asymptotics and we deduce that the asymptotic $1/p^4$ tail is approached for 
\begin{align}
\| \boldsymbol{p}\|\gtrsim p_\textrm{as} \,,\qquad p_\textrm{as} = 8k_F\,.
\label{eq:Asymptp}
\end{align} 
The Tan contact $C$ is the slope of the asymptotics and, within the present approach, its value for $T/T_F=0.9$ ($\beta\mu=-0.5$) is given by 
\begin{align}
C/k^4_F\approx 0.084 \,. 
\label{eq:TanApprox}
\end{align}
\Cref{eq:TanApprox} and further values for the Tan contact $C(\beta\mu)$ for other $\beta\mu$ are shown in \Cref{fig:contact_eos} and will be used for the following determination of the density equation of state which is an additional benchmark result: \Cref{fig:density_eos} shows the results for the normalised density $n/n_0$ as a function of dimensionless chemical potential $\beta\mu$ in comparison with other approaches. The density $n_0$ of the non-interacting Fermi gas is given by 
\begin{align}
n_0=2\int_{\bm{p}}n_F(p^2-\mu)=-2\mathrm{Li}_{3/2}(-e^{\beta\mu})/\lambda_T^3\,, 
\label{eq:n0}
\end{align}
where $\lambda_T=\sqrt{4\pi/T}$ is the thermal wavelength and $\mathrm{Li}_{3/2}$ is a poly-logarithm function~\cite{Abramowitz1972}. 

The computation of the total density \labelcref{eq:density} through the spectral function requires its large momentum tail. For an efficient computation we split the momentum integral in \labelcref{eq:density} into parts with radial momenta smaller and larger than $p_\textrm{as}$ defined in \labelcref{eq:Asymptp}, 
\begin{align}
	n =  \sum_{\pm} \Delta n^{\pm}_{p_\textrm{as}} \,,\qquad \Delta n^{\pm}_{p_\textrm{as}}= 2\int_{ \|\bm{p}\| \gtrless  p_\textrm{as}} n(\bm{p}) 
	\,.
	\label{eq:SplitDensity}
\end{align}
The low momentum part $\Delta n^-_{p_\textrm{as}}$ is computed directly from the spectral function, while we shall use the results for the asymptotic behaviour of $n(\boldsymbol{p})$ for the large momentum part $\Delta n^+_{p_\textrm{as}}$:  \Cref{eq:Tan-nas} provides us with an analytic form of the large momentum distribution and we obtain 
\begin{align}
	\Delta n^+_{p_\textrm{as}} = 2 \int_{\|\bm{p}\| >  p_\textrm{as}} \frac{C}{p^4} = \frac{C}{\pi^2 p_\textrm{as}} \,.
\end{align}
Here, we have used $p_\textrm{as} = 10 k_F$ in the explicit computation. Our real-time method reproduces the results of the well tested Luttinger-Ward approach~\cite{Haussmann:2007zz,Frank2018}. 
%

\section{Conclusion \& Outlook} 
\label{sec:Conclusion}

In the present work, we have extended the spectral functional approach \cite{Horak:2020eng} for the non-perturbative and self-consistent computation of spectral properties in non-relativistic systems. The approach was used for the direct real-time computation of fermionic single-particle and bosonic dimer spectral functions in the normal phase of a three-dimensional ultracold gas for a wide range of scattering lengths. The present results are benchmarked with observables in the unitary limit of a spin-balanced Fermi gas: the equation of state, the Tan contact and ejection rf spectra in comparison to other theoretical results and the experiment. 

The present approach opens the path towards a wide range of applications, including transport properties and the ab initio calculation of spectral functions in the superfluid phase of ultracold Fermi gases~\cite{Engelbrecht:1997zz,PhysRevB.70.094508}, and through the phase transition, including the evaluation of critical properties of the system. Its potential for computations in the broken phase of quantum systems as well as in the critical regime has already been demonstrated in relativistic systems and the respective computational advances can directly be applied here. Furthermore, the present approach can also be readily applied to general spin and mass-imbalance to study polaron physics or Fermi-Bose mixtures~\cite{vonMilczewski:2021nhz,Guo:2023spt}, where even the bosonic dimer spectral function can be measured using molecular injection spectroscopy~\cite{Diessel:2022iao}. Finally, we envisage the inclusion of vertex corrections or other classes of diagrams in systematic extensions towards full quantitative precision. We hope to report on these matters soon, but in particular on results in the superfluid phase. 

Note added: After this work was completed, two works with complementary approaches for real-time computations on the Keldysh contour were published, \cite{Johansen2023,Enss2023}, that have been triggered partially by the present work. The respective results compare well with our results, and in combination this provides a respective non-trivial systematic error control in particular within improvements. Furthermore, another independent work for the computation of self-consistent spectral functions of Fermi polarons was published very recently \cite{Hu:2023}.

\subsection*{Acknowledgements} \label{sec:Acknowledgments}

We thank M.~Bauer for providing us with lattice data for the equation of state~\cite{MarcBauer}. We thank F.~Attanasio, M.~Bauer, J.~Braun, R.~Kapust, J.~Lang, and T.~Enss  and R. ~Schmidt for discussions, and M.~W.~Zwierlein, B.~Mukherjee and B.~Frank for providing their data. E.~Dizer thanks J.~Lang also for comparison of numerical data of the present approach and Keldysh contour results. This work is funded by the Deutsche Forschungsgemeinschaft (DFG, German Research Foundation) under Germany’s Excellence Strategy EXC 2181/1 - 390900948 (the Heidelberg STRUCTURES Excellence Cluster) and the Collaborative Research Centre SFB 1225 (ISOQUANT).

\vfill

\bookmarksetup{startatroot}
\appendix

\section{Renormalisation} 
\label{app:regularization}

The four-fermi model with the action \labelcref{eq:microscopic-action} or equivalently its bosonised version with the action \labelcref{eq:BosonisedAction} requires regularisation and renormalisation. While the present model anyway has a physical cutoff given by the van der Waals length, see also the discussion below \labelcref{eq:HSrelation}, the respective standard renormalisation procedure is accommodated in the present non-perturbative spectral approach with spectral renormalisation, see \cite{Horak:2020eng, Braun:2022mgx}. 
 
In the present approximation the renormalisation procedure is easily implemented, following the standard renormalisation procedure, see \eg~\cite{PhysRevA.96.023619, PhysRevA.83.063620, Zwerger2016}. To begin with, the physical renormalisation condition can be extracted by the quantum analogue of the classical relation between the two-body scattering length $a$ in the vacuum and the full two-to-two scattering coupling at large distances, $\lambda$. In the present approximation this is simply the bubble-resummed four-fermi interaction and \labelcref{eq:lambda-a} turns into 
\begin{align} 
	-8 \pi a = h^2 G_\phi(0,\boldsymbol{0})\,.
		\label{eq:lambda-aQuantum}
\end{align}
This can be written in terms of the full physical detuning parameter 
\begin{align}
\nu =G^{-1}_\phi(0,\boldsymbol{0})= \Gamma^{(2)}_{\phi\phi^*}(0,\boldsymbol{0})\,. 
\label{eq:Quantumnu}
\end{align}
\Cref{eq:Quantumnu} is the quantum analogue of the classical relation \labelcref{eq:HSrelation} and constitutes a renormalisation condition of the dimer gap equation. A renormalisation at vanishing momentum and frequency in vacuum yields 
\begin{align}
\Pi_\phi(0,\boldsymbol{0}) =0\,,     
\label{eq:RGSigmaphi}
\end{align}
which marks the detuning parameter in the effective action \labelcref{eq:EffAct} as the physical one in \labelcref{eq:Quantumnu}. The renormalisation condition \labelcref{eq:RGSigmaphi} yields the renormalised boson DSE, 
\begin{align}\nonumber 
	G_{\phi}^{-1}(\epsilon_n,\bm{q}) &= -\frac{h^2}{8\pi a} \\[1ex] 
	& \hspace{-1cm} - h^2 \int_{\bm{p}} \left[ T\sum_{\omega_m} G_{\psi}(P) G_{\psi}(Q-P) - \frac{1}{2\bm{p}^2} \right] \,.  
	\label{eq:RenormalisedGapphi} 
\end{align}
We have used the $T$-independence of $h$ in the present approximation together with the fact, that the fermion propagator is classical in the present approximation, see~\cite{Diehl:2005an,PhysRevA.83.063620}.  In this approximation the renormalised vacuum result for $\Pi_\phi(Q)$ is readily obtained from \labelcref{eq:RenormalisedGapphi} by evaluating the self-energy loop integral with the classical fermion propagators, 
\begin{align}\nonumber 
	\Pi_{\phi}(\epsilon_n, \bm{p}) &= h^2 \int_{\bm{q}} \left[\frac{1}{-i\epsilon_n+\varepsilon_{\bm{q}}+\varepsilon_{\bm{p-q}}} - \frac{1}{2\varepsilon_{\bm{q}}}\right]  \\[2ex]
	&= -\frac{h^2}{8\pi} \sqrt{-\frac{i\epsilon_n}{2}+\frac{\bm{p}^2}{4}}  \,, 
\end{align}
for general complex frequencies $\epsilon_n\in\mathbbm{C}$. This concludes our discussion of the renormalisation of the bosonised theory within the present approximation. 

\newpage

\section{Boson self-energy calculation} \label{app:boson-self-energy-calculation}

In this Appendix, we discuss the explicit computations and analytic results for the boson self-energy. Starting from~\labelcref{eq:spectral-boson-self-energy}, we define
\begin{align}
	\Pi_{\phi}(\epsilon_n,\bm{p}) = h^2\int_{\bm{q}}\left[\int_{\lambda_1,\lambda_2} \rho_{\psi}(\lambda_1,\bm{q}) \rho_{\psi}(\lambda_2,\bm{p}-\bm{q}) \right. \notag \\[1ex]
	\left. \times\, I(\epsilon_n,\lambda_1,\lambda_2) - \frac{1}{2\bm{q}^2} \right] \,,
\end{align}
with the analytic sum over fermionic Matsubara frequencies $\omega_m=(2m+1)\pi T$, 
\begin{align}
	I(\epsilon_n,\lambda_1,\lambda_2) &= T\sum_{\omega_m} \frac{1}{i\omega_m-\lambda_1}\frac{1}{i(\epsilon_n-\omega_m)-\lambda_2} \notag \\
	&= \frac{1-n_F(\lambda_1)-n_F(\lambda_2)}{-i\epsilon_n+\lambda_1+\lambda_2} \,,
\end{align}
Performing the analytic continuation $i\epsilon_n\rightarrow\omega+i0^+$ and taking the imaginary part, we end up with~\Cref{eq:imaginary-part-boson-self-energy}.

For the first iteration, analytic expressions for the self-energy at finite and zero temperature can be derived. Inserting the classical fermion spectral function $\rho_{\psi}(\lambda,\bm{p})=\delta(\lambda-\bm{p}^2+\mu)$, we obtain the well-known expression for the retarded boson self-energy~\cite{Nozieres:1985zz, Punk2010}
\begin{align}\nonumber 
	\Pi^R_{\phi}(\omega, \bm{p}) = & h^2 \int_{\bm{q}} \Biggl[ \frac{1-n_F(\varepsilon_{\bm{q}}-\mu) -n_F(\varepsilon_{\bm{p-q}}-\mu)}{-\omega+\varepsilon_{\bm{q}} + \varepsilon_{\bm{p-q}}-2\mu-i0^+}\\[1ex]
	& \hspace{1cm} - \frac{1}{2\varepsilon_{\bm{q}}} \Biggl] \,,
\end{align}
where $\varepsilon_{\bm{p}}=\bm{p}^2$ is the classical momentum dispersion. The boson self-energy can be separated in a temperature-independent and a temperature-dependent part, 
\begin{align}
\Pi^R_{\phi} = \Pi^{R,0}_{\phi} + \Pi^{R,T}_{\phi}\,.
\end{align}
After the shift $\bm{q} \rightarrow \bm{q} + \bm{p}/2$, the angular integration is trivial, and we obtain the finite expression for the vacuum part,
\begin{align}\nonumber 
\Pi^{R,0}_{\phi}(\omega, \bm{p}) &= \int_{\bm{q}} \left[\frac{h^2 }{-\omega+\varepsilon_{\bm{q}}+\varepsilon_{\bm{p-q}}-2\mu-i0^+} - \frac{h^2 }{2\varepsilon_{\bm{q}}}\right] \\[1ex]
&= -\frac{h^2}{8\pi} \sqrt{-y-i0^+}  \,,
\end{align}
where we have defined $y = \omega/2-p^2/4+\mu$. The temperature-dependent part can be obtained analogously. We note that the contribution from both Fermi distributions is identical in the spin-balanced case and find
\begin{align}\nonumber 
	\Pi^{R,T}_{\phi}(\omega, \bm{p}) &= -h^2 \int_{\bm{q}}
	\frac{n_F(\varepsilon_{\bm{q}}-\mu)+n_F(\varepsilon_{\bm{p-q}}-\mu)}{-\omega+\varepsilon_{\bm{q}} + \varepsilon_{\bm{p-q}}-2\mu-i0^+} \\[2ex]
		&= -\frac{h^2}{4\pi^2} \int_0^{\infty} \frac{2\chi(q) \, q^2 \, dq}{q^2-y-i0^+} \,,
\end{align}
with the angle-integrated function
\begin{widetext} 
\begin{align} 
	\chi(q) = \int_{-1}^1 \frac{dx}{2} \,
	n_F\left(\left[\bm{q}\pm\bm{p}/2\right]^2-\mu\right) =
	\begin{cases}
		n_F\left(q^2-\mu\right) & \quad \textrm{for} \quad p=0 \\[2ex]
		\frac{T}{2pq} \ln\left[\frac{
			n_F\left(\mu-q^2_{+}\right)}{
			n_F\left(\mu-q^2_{-}\right)}
		\right] & \quad \textrm{for} \quad p \neq 0
	\end{cases} \,,
\end{align}
with $q_{\pm}=q\pm p/2$.
Here, $x=\cos(\theta_{\bm{pq}})$ and $\theta_{\bm{pq}}$ is the angle between the vectors $\bm{p}$ and $\bm{q}$. Note that this function yields a non-zero contribution in the vacuum for $\mu>0$. In the limit of $T\rightarrow 0$, the Fermi functions $n_F(x)\rightarrow\theta(-x)$, and the expression simplifies to
\begin{align} 
	\chi^{T=0}(q) =
	\begin{cases}
		\theta\left(\mu-q^2\right) & \quad \textrm{for} \quad  p = 0 \\[2ex]
		\frac{\theta\left(\mu-q^2_{-}\right)}{2pq} \left( \mu-q^2_{-} - (\mu-q^2_{+}) \theta\left(\mu-q^2_{+}\right)
		\right) &\quad \textrm{for} \quad p \neq 0
	\end{cases} \,.
\end{align}
The real and imaginary part can be obtained analytically by using 
\begin{align}
	\frac{1}{x-i0^+} = P\left(\frac{1}{x}\right) + i\pi\delta(x) \,.
\end{align}
There are the following two cases:
If $y<0$, the integrals for the real part are well-defined and $\mathrm{Im}\,\Pi^{R,T}_{\phi} = 0$.
If $y\geq 0$, the imaginary part is given analytically by
\begin{align}
	\mathrm{Im}\,\Pi^{R,T}_{\phi}(\omega, \bm{p}) = -\frac{h^2}{4\pi}\int_0^{\infty}dq\,2\chi(q)q^2\delta(q^2-y)  = -\frac{1}{4\pi}\sqrt{y}\chi(\sqrt{y}) \,,
\end{align}
and the real part $\mathrm{Re}\,\Pi^{R,T}_{\phi}$ can be obtained numerically via a one-dimensional principal value integral, see~\cite{PhysRevA.105.043303}.

For completeness, we also give the analytic expressions for the real part at $T=0$. Since the contribution from both Fermi functions is identical, we obtain for $\text{Re}\,\Pi^{R,T=0}_{\phi}$ at zero temperature ($\mu > 0$), 
\begin{align}
	\text{Re}\,\Pi^{R,T=0}_{\phi}(\omega, \bm{p}) =\,
	\frac{h^2}{4\pi^2} \left[
	\sqrt{\mu} - \frac{y-\mu+\frac{p^2}{4}}{2p}\log\left(\frac{y-\xi^2_{+}}{y-\xi^2_{-}}\right)
	- \sqrt{|y|}
	\begin{cases}
		\text{arctanh}\left(\frac{\xi_{-}}{\sqrt{y}}\right) + 
		\text{arctanh}\left(\frac{\xi_{+}}{\sqrt{y}}\right)  & \quad \textrm{for} \quad y \geq 0 \\[2ex]
		\text{arctan}\left(\frac{\xi_{-}}{\sqrt{|y|}}\right) + 
		\text{arctan}\left(\frac{\xi_{+}}{\sqrt{|y|}}\right) & \quad \textrm{for} \quad  y < 0
	\end{cases}
	\right] \,,
\end{align}
with $\xi_{\pm} = \sqrt{\mu} \pm p/2$. The $p=0$ limit is given by
\begin{align}
	\text{Re}\,\Pi^{R,T=0}_{\phi}(\omega, \bm{0}) &= \frac{h^2}{2\pi^2} 
	\begin{cases}
		\sqrt{\mu}-\sqrt{y}\,
		\text{arctanh}\left(\sqrt{\frac{\mu}{y}}\right)  & \quad \textrm{for} \quad y \geq 0 \\[2ex]
		\sqrt{\mu} - \sqrt{|y|} \arctan\left( \sqrt{\frac{\mu}{|y|}}\right)  & \quad \textrm{for} \quad y < 0
	\end{cases}	\,.
\end{align}
\begin{figure}[t]
	\subfloat[$\mathrm{Re}\,\Pi_{\phi}^R(\omega,\bm{p})/\sqrt{\varepsilon_F}$]{\includegraphics[width=0.49\linewidth]{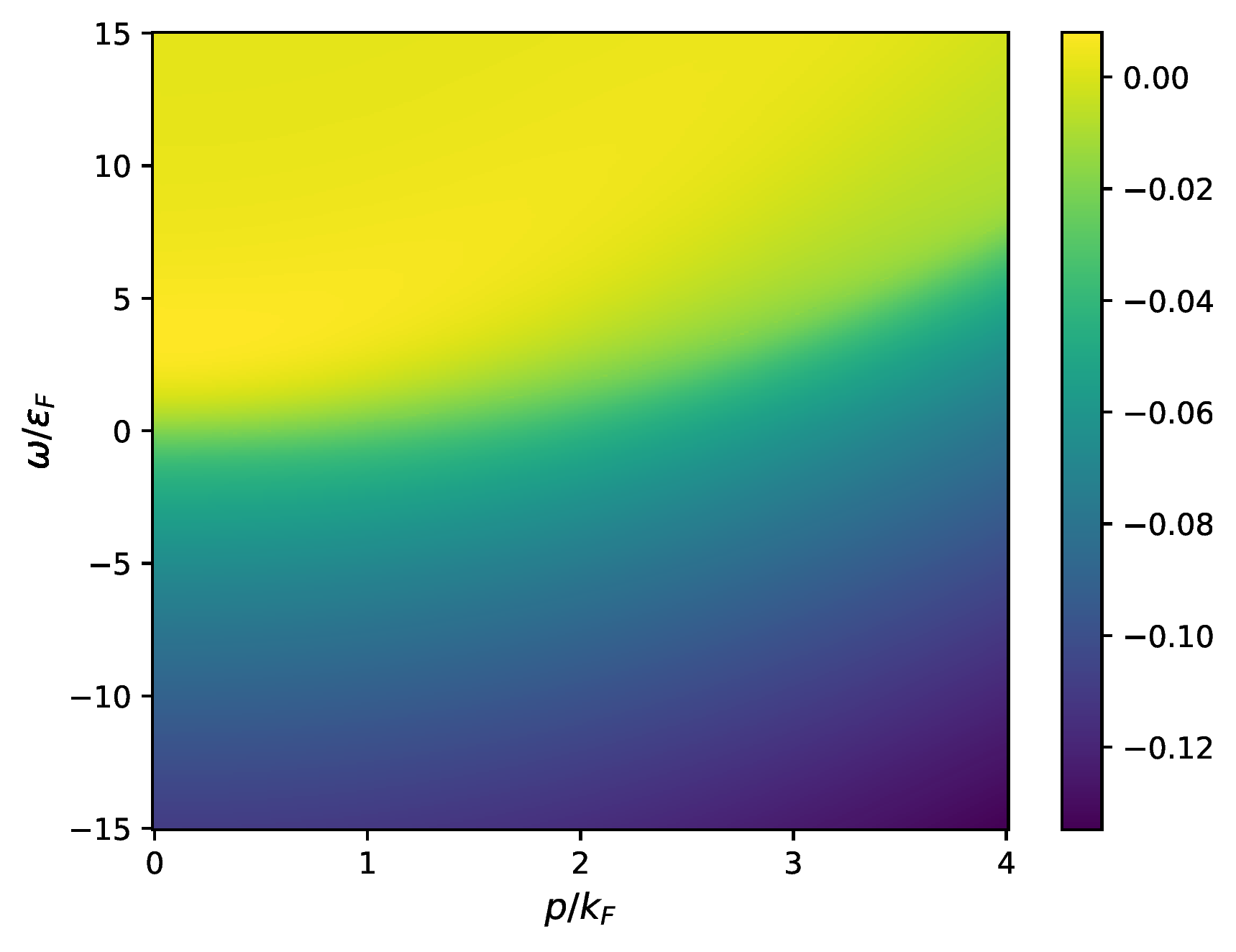}}
	\subfloat[$\mathrm{Im}\,\Pi_{\phi}^R(\omega,\bm{p})/\sqrt{\varepsilon_F}$]{\includegraphics[width=0.48\linewidth]{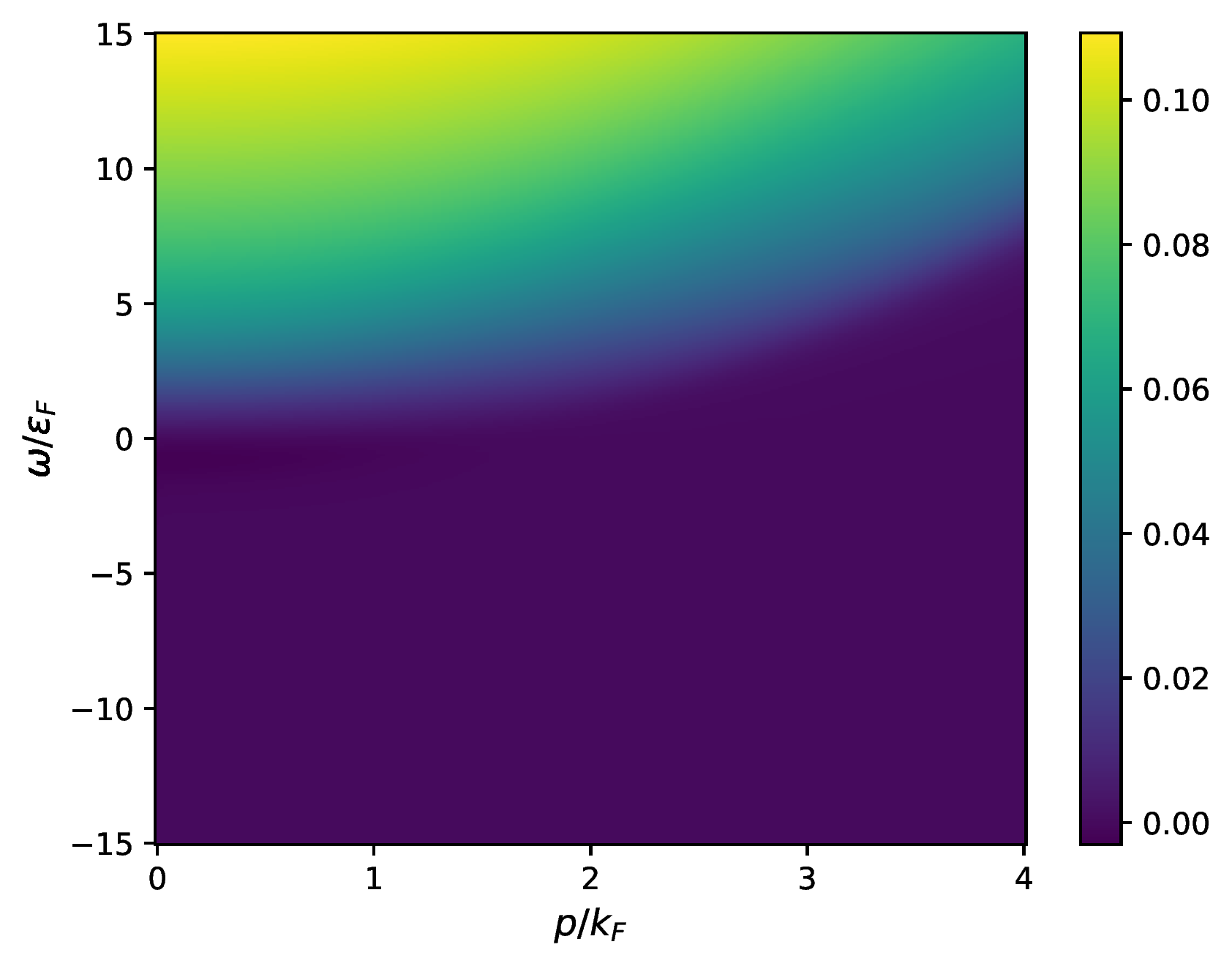}}
	\caption{Real and imaginary part of the retarded boson self-energy $\Pi_{\phi}^R(\omega,\bm{p})$ for the balanced Fermi gas at unitarity and $T/T_F=0.56$ ($\beta\mu=0.13$).}
	\label{fig:boson-self-energy}
\end{figure}
The contribution from the next iterations has to be computed numerically and gives rise to a finite correction on top of the analytic result. For more details on the numerical treatment, see~\Cref{app:implementation}. Finally, an exemplary plot of the fully converged boson self-energy $\Pi_{\phi}^R(\omega,\bm{p})$ is shown in~\Cref{fig:boson-self-energy}.

\section{Fermion self-energy calculation} \label{app:fermion-self-energy-calculation}

For the fermion self-energy, the Matsubara sum can also be calculated analytically. However, there is no analytic result for the first iteration since the classical boson spectral function is not defined. Therefore the fermion self-energy has to be computed numerically. Starting from~\Cref{eq:spectral-fermion-self-energy}, we define
\begin{align}
	\Sigma_{\psi}(\omega_n,\bm{p}) = \int\limits_{\lambda_1,\lambda_2,\bm{q}}  \hspace{-.3cm}\rho_{\phi}(\lambda_1,\bm{q}) \rho_{\psi}(\lambda_2,\bm{p}-\bm{q}) I(\omega_n,\lambda_1,\lambda_2) \,,
\end{align}
with the analytic Matsubara sum over bosonic frequencies $\epsilon_m=2m\pi T$, 
\begin{align}
	I(\omega_n,\lambda_1,\lambda_2) = T\sum_{\epsilon_m} \frac{1}{i\epsilon_m-\lambda_1}\frac{1}{i(\epsilon_m-\omega_n)-\lambda_2}= \frac{-n_B(\lambda_1)-n_F(\lambda_2)}{-i\omega_n+\lambda_1-\lambda_2} \,.
\end{align}
Performing the analytic continuation $i\omega_n\rightarrow\omega+i0^+$ and taking the imaginary part, we recover~\Cref{eq:imaginary-part-fermion-self-energy}. Note that the pole in $n_B(\lambda_1)$ at $\lambda_1=0$ is exactly cancelled by the zero-transition of $\rho_{\phi}(\lambda_1)$.

For the numerical computation of the fermion self-energy, the high-frequency tails contribute significantly. Therefore, we adopt a semi-analytic treatment which is discussed in the next section and improves the determination of the real part via Kramers-Kronig substantially. An exemplary plot of the fully converged fermion self-energy $\Sigma_{\psi}^R(\omega,\bm{p})$ is shown in~\Cref{fig:fermion-self-energy}.

\begin{figure}[t]
	\subfloat[$\mathrm{Re}\,\Sigma_{\psi}^R(\omega,\bm{p})/\varepsilon_F$]{\includegraphics[width=0.49\linewidth]{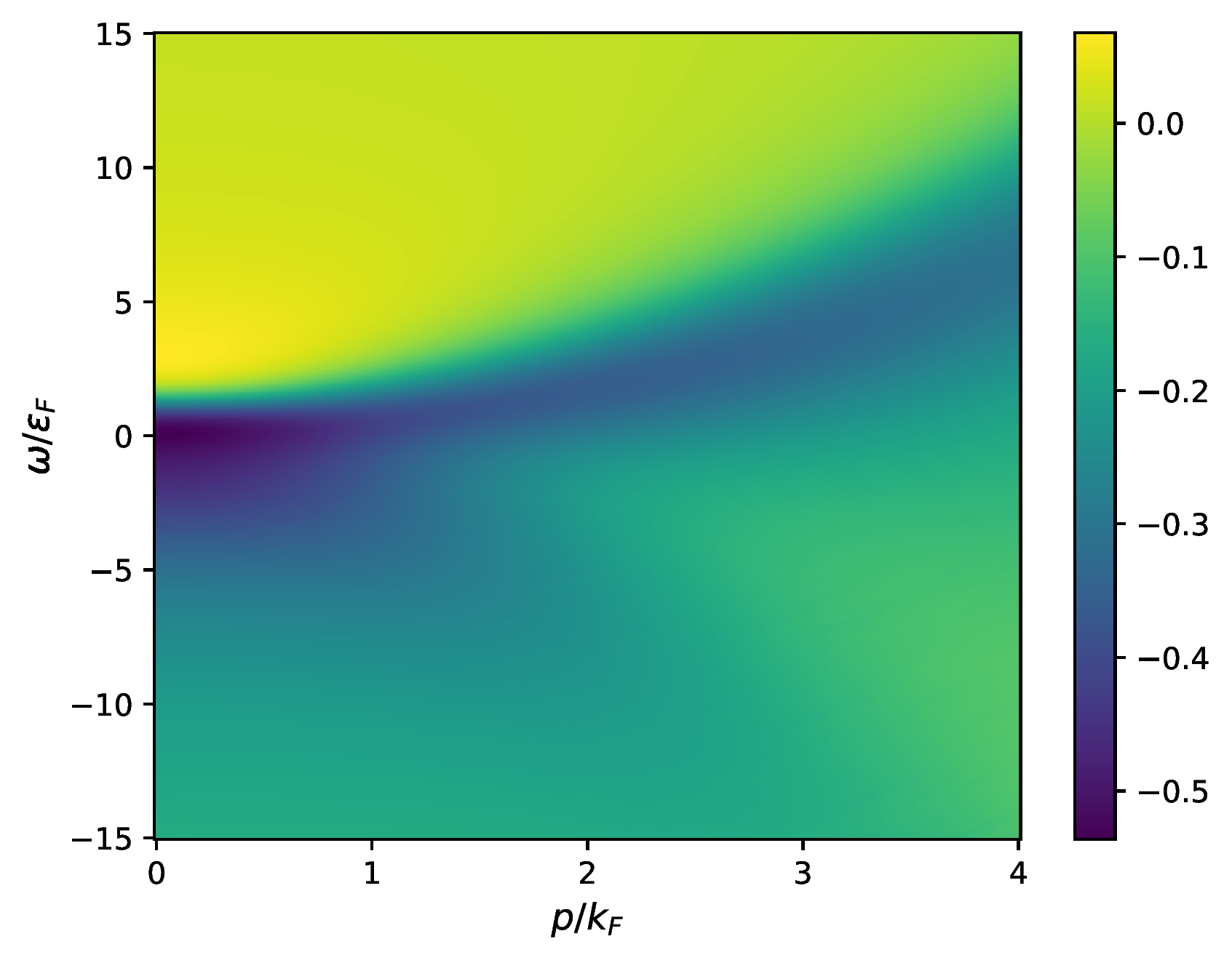}}
	\subfloat[$\mathrm{Im}\,\Sigma_{\psi}^R(\omega,\bm{p})/\varepsilon_F$]{\includegraphics[width=0.49\linewidth]{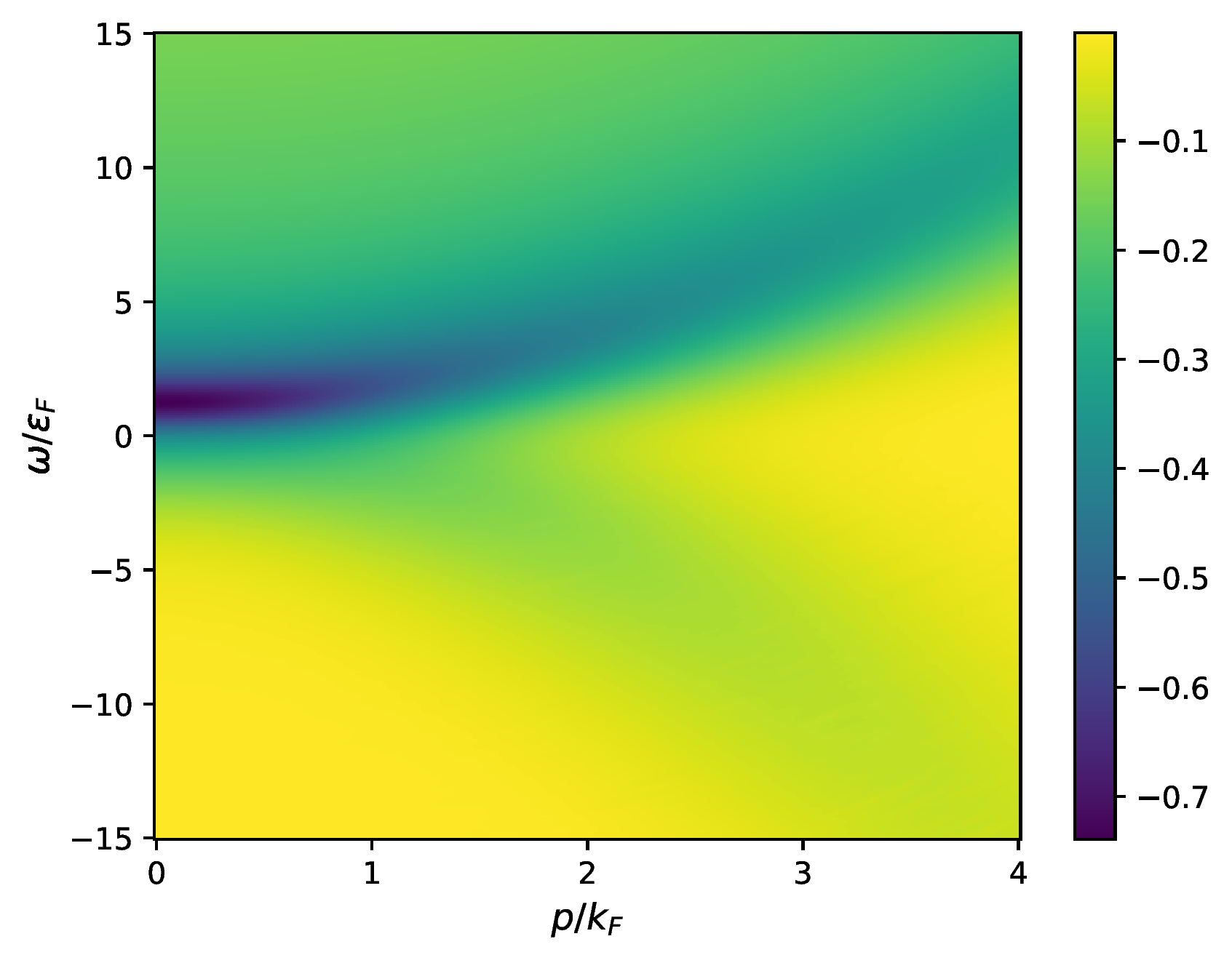}}
	\caption{Real and imaginary part of the retarded fermion self-energy $\Sigma_{\psi}^R(\omega,\bm{p})$ for the balanced Fermi gas at unitarity and $T/T_F=0.56$ ($\beta\mu=0.13$).}
	\label{fig:fermion-self-energy}
\end{figure}

\end{widetext} 
%

\section{Numerical implementation} \label{app:implementation}

The numerical iteration is done with a standard iteration procedure for integral equations. The first iteration step is done for the bosonic propagator and only involves the classical fermion propagators. The following one for the fermion propagator includes the results of the first step for the bosonic propagator and again uses the classical fermion propagator. All the remaining ones use the fully numerical results from the previous steps. The respective workflow can be summarised as follows, 
\begin{enumerate}

\item \textit{First iteration of the boson propagator: } Start with the analytic expression for $\text{Im}\,\Pi^R_{\phi}$ and calculate $\text{Re}\,\Pi^R_{\phi}$ on a finite grid. To simplify the calculation, treat the divergent vacuum part analytically, see below.
Choose grid size large enough such that the numerical corrections are small. This can be quantified in dependence of the temperature.

\item \textit{First iteration of the fermion propagator: } Take the classical fermion spectral function (delta peak) and the semi-analytic boson spectral function from the first iteration and calculate $\text{Im}\,\Sigma^R_{\psi}$ with~\labelcref{eq:imaginary-part-fermion-self-energy} on a finite grid. From this, obtain $\text{Re}\,\Sigma^R_{\psi}$ as described below.

\item  \textit{Further iterations of the fermion propagator: } Take the numerical spectral functions for the fermion and boson and calculate the self-energy as in step 2.

\item  \textit{Further iterations of the boson propagator: } Take numerical spectral functions for the fermions and calculate $\text{Im}\,\Pi^R_{\phi}$ on a finite grid. Compute the difference to the non-self-consistent result and obtain the real part. Outside the grid, glue smoothly to the semi-analytical non-self-consistent result.

\end{enumerate}
For the numerical calculation of the two-dimensional functions an adaptive method is used. The sampling points are chosen automatically based on the functional form. Less samples are taken in slow varying regions and more samples are taken in faster varying regions. Additionally, the calculation of these sampling points can be parallelised over multiple cores. The resulting sampling points are then linearly interpolated. It is useful to simplify the interpolation by transforming the functions onto the quadratic dispersion relation. For the bosonic self-energy, this step can significantly improve the resolution of sharp edges in the function. For the fermionic self-energy, it can additionally help to ensure a large enough distance of the grid boundary to the main peak, such that asymptotic behaviour is guaranteed for all momenta. 

The numerical integration of the three-dimensional self-energy loop integrals is performed mainly with an adaptive Monte Carlo method. This allows for maximal flexibility when dealing with highly peaked integrands in a multidimensional space and can be generalised easily. However, it comes with the downside of a long runtime in comparison to standard adaptive routines in low dimensions like here. For this specific case, a different adaptive integration routine using sparse grids is far more efficient. The one-dimensional principal value integral for the real part of the self-energy is computed efficiently using an adaptive quadrature integration.

For the representation of the numerical self-energy on the finite grid, an adaptive grid is chosen for the imaginary and real part, separately. Typical boundaries in frequency are $\omega=[-200,200]\,\varepsilon_F$ and in momentum $p=[0,10]\,k_F$. Approximately 10.000-20.000 grid points are needed to obtain stable numerical results.

\subsection{Boson spectral function} \label{subsec:boson_spec}

In this Section, we detail the numerical calculation and representation of the boson spectral function $\rho_{\phi}$. Since the vacuum part of the bosonic self-energy is problematic, the numerical procedure requires suitable subtraction schemes and analytic treatment. 

As seen above, the vacuum contribution to the boson self-energy is given by 
\begin{align}
	\Pi^{0}_{\phi}(\epsilon_n, \bm{p}) = -\frac{h^2}{8\pi} \sqrt{-\frac{i\epsilon_n}{2}+\frac{\bm{p}^2}{4}-\mu}  \,, 
	\label{eq:Phi0Gen}
\end{align}
With \labelcref{eq:Phi0Gen}, the (retarded) imaginary part follows as 
\begin{align}
	\mathrm{Im}\,\Pi^{R,0}_{\phi}(\omega, \bm{p}) = \frac{h^2}{8\pi} \sqrt{\frac{\omega}{2}-\frac{\bm{p}^2}{4}+\mu} \,. 
	\label{eq:Phi0Ret}
	\end{align}
In order to obtain the full real part $\mathrm{Re}\,\Pi^{R}_{\phi}$ of the self-energy via Kramers-Kronig numerically, we can subtract \Cref{eq:Phi0Ret} from the total numerical imaginary part $\mathrm{Im}\,\Pi^{R}_{\phi}$ and find
\begin{align}\nonumber 
	\mathrm{Re}\,\Pi^{R}_{\phi}(\omega, \bm{p}) &= \mathrm{Re}\,\Pi^{R,0}_{\phi}(\omega, \bm{p}) \\[1ex]
	&+ \frac{1}{\pi} \int_{\lambda} \frac{\mathrm{Im}\,\Pi^R_{\phi}(\lambda,\bm{p})-\mathrm{Im}\,\Pi^{R,0}_{\phi}(\lambda,\bm{p})}{\lambda-\omega} \,,
\end{align}
where the real part of the vacuum solution is
\begin{align}
	\mathrm{Re}\,\Pi^{R,0}_{\phi}(\omega, \bm{p}) = -\frac{h^2}{8\pi} \sqrt{-\frac{\omega}{2}+\frac{\bm{p}^2}{4}-\mu} \,.
\end{align}
With these formulas, we can obtain $\mathrm{Im}\,\Pi^{R}_{\phi}$ and $\mathrm{Re}\,\Pi^{R}_{\phi}$ on a finite grid numerically. However, the calculation of the real part requires also information about the high frequency tails outside the numerical grid. In this case, the imaginary part outside the grid is approximated by the analytic formula of the non-self-consistent self-energy discussed in the previous Section,
\begin{align}
	\mathrm{Im}\,\Pi^{R}_{\phi}(\omega, \bm{p})-\mathrm{Im}\,\Pi^{R,0}_{\phi}(\omega, \bm{p}) \approx \mathrm{Im}\,\Pi^{R,T}_{\phi}(\omega, \bm{p}) \,,
\end{align}
for large $\omega$. Thus, the bosonic spectral function outside the grid is approximated by the non-self-consistent (first iteration) spectral function $\rho^{(1)}_{\phi}$. A different subtraction scheme would be to subtract the whole non-self-consistent imaginary part straight away and only deal with differences to the first iteration. In principle, both methods are equivalent and work similar. For practical reasons, we choose the latter subtraction scheme.

Another trick to improve the numerical calculation was already mentioned above. Since the boson self-energy follows a quadratic dispersion relation, it is practical to transform the functions onto the dispersion relation before interpolation. This way, a better sampling and interpolation of the important regions can be achieved. Afterwards, the interpolated function is shifted back to the correct dispersion relation.

Since the fermion spectral functions in the self-energy integrals are very peaked for lower temperatures and larger momenta, it might be useful to subtract a broadened classical spectral function, or the first iteration, from the further iterations in order to improve the integration for larger momenta. The contribution from the subtracted spectral function has then to be taken into account. Since the fermion spectral functions are represented on a finite numerical grid, the bare delta peak contributions from outside the grid has to be taken into account too. This is similar to the peak-tail splits from~\cite{Horak:2020eng}.

\subsection{Calculation of the real part via Kramers-Kronig} \label{subsec:real-part}

It turns out that the calculation of the real part via Kramers-Kronig relation is very sensitive to the high frequency tails of the imaginary part. We accommodate for this fact with an extrapolation of the high frequency tail, which allows us to estimate its contribution. For constant momentum $p$, the large-frequency asymptotics of the fermion self-energy is captured by the power law
\begin{align}
	\lim_{\omega\rightarrow\infty} \mathrm{Im}\,\Sigma^{R}_{\psi}(\omega) = a\, \omega^{-1/2} \,,
\end{align}
where the constant $a$ is determined by a fitting routine. From this fit one can calculate the missing contribution for the real part at a frequency $\omega$ via
\begin{align}
	\delta\mathrm{Re}\,\Sigma^{R}_{\psi}(\omega) = \frac{2a}{\pi\sqrt{\Delta}}\,{}_{2}F_{1}\left(\frac{1}{2},\frac{1}{2};\frac{3}{2};-\frac{\omega}{\Delta}\right) \,,
\end{align}
where $\Delta=\omega_{\mathrm{max}}-\omega$ and $\omega_{\mathrm{max}}$ is the largest frequency of the grid, and ${}_2F_1(a,b;c;z)$ being the hypergeometric function~\cite{Abramowitz1972}. Using this large-frequency contribution from outside the grid, improves the determination of the fermion real part significantly.

This asymptotic behaviour can be shown by similar arguments as in Ref.~\cite{Schneider2009}. We consider the large frequency behaviour and rewrite the imaginary part of the fermion self-energy~\labelcref{eq:imaginary-part-fermion-self-energy} as
\begin{align}
	\mathrm{Im}\,\Sigma^R_{\psi}(\omega) \sim \int_{\lambda_1,\lambda_2,\bm{q}} &\rho_{\phi}(\lambda_1) \rho_{\psi}(\lambda_2) \delta(\omega-\lambda_1+\lambda_2) \notag \\[1ex]
	&\times \left[n_B(\lambda_1)+n_F(\lambda_2)\right] \,.
\end{align}
In the limit $\omega\rightarrow\infty$, the delta function contributes only if (a) $\lambda_1$ large and $\lambda_2$ small or (b) $\lambda_2$ large negative and $\lambda_1$ small. In case (b), however, the fermion spectral function $\rho_{\psi}$ vanishes for small momenta and $\lambda_2\rightarrow-\infty$. Thus, only case (a) contributes and we are left with
\begin{align}
	\mathrm{Im}\,\Sigma^R_{\psi}(\omega) \sim \int_{\lambda_2,\bm{q}} \rho_{\phi}(\omega+\lambda_2) \rho_{\psi}(\lambda_2) n_F(\lambda_2) \,,
\end{align}
where we have used $n_B(\lambda_1)\rightarrow 0$ for $\lambda_1\rightarrow\infty$. As seen above, the large frequency behaviour of the boson spectral function is dominated by $\rho_{\phi}(\omega)\sim 1/\sqrt{\omega}$. Additionally, the integrand vanishes for large negative values of $\lambda_2$, since $\rho_{\psi}(\lambda_2)$ vanishes, and for large positive values of $\lambda_2$, since $n_F(\lambda_2)$ vanishes. Thus, the range of $\lambda_2$ is limited and we can write
\begin{align}
	\mathrm{Im}\,\Sigma^R_{\psi}(\omega) \sim \int_{\lambda_2,\bm{q}} \rho_{\psi}(\lambda_2) n_F(\lambda_2) / \sqrt{\omega} \sim 1/\sqrt{\omega} \,,
\end{align}
where we used that $\int_{\lambda_2,\bm{q}} \rho_{\psi}(\lambda_2) n_F(\lambda_2) = n$ is finite in the last step. With the same argument one can show that $\mathrm{Im}\,\Sigma^R_{\psi}(\omega)$ is mostly suppressed for large negative frequencies.

\subsection{Iterative procedure} \label{subsec:iteration}

Using~\labelcref{eq:imaginary-part-fermion-self-energy} and~\labelcref{eq:imaginary-part-boson-self-energy}, and the Kramers-Kronig relation, the retarded self-energies can be computed in every iteration step and fed back into the next, after extraction of the spectral function via~\Cref{eq:spectral-relation}. The iterative procedure is initialised with the classical fermion spectral function,
\begin{align} \label{eq:inital-fermion-spec}
	\rho^{(0)}_{\psi}(\omega,\bm{p}) = \delta(\omega-\bm{p}^2+\mu) \,.
\end{align}
This initial guess is inserted into the spectral form of $\Pi_{\phi}(P)$ to obtain the first iteration of the boson spectral function $\rho_{\phi}^{(1)}$. Then, $\rho_{\phi}^{(1)}$ together with $\rho^{(0)}_{\psi}$ are inserted into the spectral integral of $\Sigma_{\psi}(P)$ to obtain the first iteration of the fermion spectral function $\rho_{\psi}^{(1)}$. Now, $\rho_{\psi}^{(1)}$ can be used to obtain the next iteration for the boson spectral function $\rho_{\phi}^{(2)}$, and so on. In general, $\rho_{\psi}^{(i)}$ is used to obtain $\rho_{\phi}^{(i+1)}$, and $\rho_{\phi}^{(i+1)}$ and $\rho_{\psi}^{(i)}$ are used to obtain $\rho_{\psi}^{(i+1)}$. This iteration is repeated until convergence is reached. We observe that around 5-20 iterations are needed to obtain a converged result with
\begin{align}
\int_{\lambda} \left\| \rho^{(i)}(\lambda,\bm{p})-\rho^{(i-1)}(\lambda,\bm{p}) \right\| \lesssim 0.005 \,, \quad \forall \bm{p} \,.
\end{align}
As mentioned in the main text, the convergence is worse closer to the critical temperature and the spectral functions may oscillate between intermediate solutions. One can improve the convergence by iterating twice over the fermions or updating the spectral functions only partially.

\bibliography{bib}

\end{document}